\documentclass[a4paper,aps,prd,twocolumn,showpacs,eqsecnum,amsmath,amssymb,nofootinbib,superscriptaddress]{revtex4-1}
\usepackage{dcolumn}
\usepackage{bm}
\usepackage{graphics}
\usepackage{graphicx}
\usepackage{epsfig}
\usepackage{booktabs,multirow}
\usepackage{hhline}
\usepackage{slashed}
\usepackage{rccol}
\usepackage{mathrsfs}
\usepackage{xcolor,colortbl}

\usepackage{xcolor}\colorlet{linkequation}{blue}
\usepackage[colorlinks=true, 
            linkcolor=blue,   
            filecolor=magenta,   
            citecolor=magenta, 
            urlcolor=cyan,  
            hyperfootnotes=true]{hyperref}
\usepackage{footnotebackref}

\newcommand*{\SavedEqref}{}
\let\SavedEqref\eqref
\renewcommand*{\eqref}[1]{%
  \begingroup
    \hypersetup{
     linkcolor=linkequation,
      linkbordercolor=linkequation,
    }%
    \SavedEqref{#1}%
  \endgroup

}

\begin{document}


\def\beqa{\begin{eqnarray}}
\def\eeqa{\end{eqnarray}}
\newcommand{\be}{\ensuremath{\beta}}
\newcommand{\al}{\ensuremath{\alpha}}
\newcommand{\sa}{\ensuremath{\sin\alpha}}
\newcommand{\ca}{\ensuremath{\cos\alpha}}
\newcommand{\ta}{\ensuremath{\tan\alpha}}
\newcommand{\sbt}{\ensuremath{\sin\beta}}
\newcommand{\cbt}{\ensuremath{\cos\beta}}
\newcommand{\ma}{\ensuremath{m_{A}}}

\newcommand{\ben}{\begin{enumerate}}
\newcommand{\een}{\end{enumerate}}
\newcommand{\bc}{\begin{center}}
\newcommand{\ec}{\end{center}}
\newcommand{\mb}{\mbox{\ }}
\newcommand{\vs}{\vspace}
\newcommand{\ra}{\rightarrow}
\newcommand{\la}{\leftarrow}
\newcommand{\ul}{\underline}
\newcommand{\ds}{\displaystyle}
\definecolor{LightCyan}{rgb}{0.0, 1, 0.94}

\title{Pseudoscalar Higgs boson pair production at a photon-photon collision in the two Higgs doublet model}

\author{M.~Demirci}
\email{mehmetdemirci@ktu.edu.tr}
\affiliation{Department of Physics, Karadeniz Technical University, 61080 Trabzon, Turkey}%
\date{\today}
\begin{abstract} In this study, the direct pair production of the pseudoscalar Higgs boson at a photon-photon collision is analyzed in the context of two Higgs doublet model, taking account the complete one-loop contributions. In order to illustrate the effect of the new physics, four benchmark points scenarios, which are consistent with theoretical and current experimental constraints, are chosen in the type-I of THDM with an exact alignment limit. In these benchmark points, the CP even lightest Higgs boson ($h^0$) have the Standard Model like couplings to the gauge bosons. The effect of individual contributions from each type of one-loop diagrams on the total cross section are examined in detail. The dependence of total cross section on the center-of-mass energy is also presented at the various polarization configurations of the incoming photons. Moreover, the regions $m_{12}^2-\tan\beta$ and $m_A-\tan\beta$ in the parameter space of the THDM are scanned for some fixed values of other parameters. The box-type diagrams make a much larger contribution to the total cross section than the others at high energies. Total cross section can be enhanced by a two factor thanks to opposite polarized photons as well as threshold effects.
\keywords{Two Higgs doublet model, pseudoscalar Higgs boson, photon-photon collider, future linear collider}
\end{abstract}

\maketitle

\section{Introduction}
One of the most important achievements of the Large Hadron Collider (LHC) has been the discovery
of a resonance about 125 GeV\footnote{The combined mass measurement obtained from the data at $\sqrt{s}=$ 7 and 8 TeV by the ATLAS and CMS experiments is $m_h=125.09 \pm 0.21(\text{stat}.)\pm 0.11(\text{syst}.)$ GeV.}~\cite{ATLAS,CMS}, whose measured signal rates in dominant decay channels increasingly comply with that of the Higgs boson of Standard Model (SM)~\cite{ATLASCMS}. However, there are still mysteries here. The Higgs couplings are not universal, as the gauge couplings are, and their pattern is not explained by the SM. The observation of a Higgs boson further opens the door for the possibility of extended Higgs
sectors, with parameters constrained by measured properties. The Two Higgs Doublet Model (THDM)~\cite{THDM1} is one of the simplest such extensions, which introduces one additional
Higgs doublet. Versions of the THDM appear in various of well-motivated scenarios
for new physics beyond SM, both with and without supersymmetry (SUSY)~\cite{Haber1985}, where the extra Higgs doublet is either a necessary component or essential by-product in addressing problems such as the gauge hierarchy problem, the
origin of dark matter, the strong CP problem and the generation of a baryon asymmetry. Additionally, given the multiplicity of Higgs states in a THDM, its scalar potential is significantly more involved than the SM one. It has a multitude of triple self-couplings, unlike the SM, which only has one. Such interactions are key to understanding the phenomenology of the THDM, because they identy the form of the potential. To test of the nature of the Higgs bosons, particularly it will be very important to the measurement of the such couplings of the Higgs bosons to the other particles. The measurement at the LHC is rather challenging, due to requiring huge luminosity. In this respect, future linear colliders will play an important role: The clean environment in these colliders will ensure that these couplings are precisely identified as model-independent. One of most advanced design for a future lepton collider is the International Linear Collider (ILC)~\cite{ILC1,ILC2} which is designed to give facilities for $e^-e^+$ along with other options such as $e^-e^-$, $e^-\gamma$ and $\gamma\gamma$ collisions. The positron and electron beams are foreseen to be polarised to $\pm 30\%$ and $\pm 80\%$, respectively.
Also, there is an organisation that brings ILC and the Compact Linear Collider (CLIC)~\cite{CLILC1} projects together under one roof, which is called the Linear Collider Collaboration (LCC)~\cite{LCC}. The primary task of the LCC will be to extend and complement the results obtained at the LHC, and to explore new physics beyond the SM. The $\gamma\gamma$-collider is also considered as a next option with an integrated luminosity of the order of 100 fb$^{-1}$ yearly. The machine is expected to be upgradeable to the center-of-mass energy range of $\sqrt{s} = 1000$ GeV with a total integrated luminosity up to 300 fb$^{-1}$ yearly~\cite{ILC3}. Besides the possibility to discover relatives of the Higgs boson via studying the properties of the 125 GeV Higgs boson, the ILC provides excellent opportunities to discover additional lighter Higgs bosons –or, more generally, any weakly interacting light scalar or pseudo-scalar particle– by their direct production~\cite{Fujii2017}.

The main mechanism of production pseudoscalar Higgs boson at a $\gamma \gamma$ collider is $\gamma \gamma \to A^0$
\cite{Gunion1,spira,Asner}, however in order to research the relevant quartic and triple couplings at future linear colliders, the pair production mode is necessary to be studied. Additionally, a $\gamma\gamma$-collider will supply a distinct way to produce the pseudoscalar Higgs boson pair which deserves a detailed study. Furthermore, production of neutral particle pairs in photon-photon collisions may be significantly sensitive to new physics effects as a such process is naturally subdued since they first emerge at the one-loop level, thereby providing a detailed test for the structure of extended Higgs sectors. The triple Higgs couplings in the THDM were widely examined at electron-positron linear colliders~\cite{linear} and shown to supply an opportunity  for measurement of those couplings. The pseudoscalar Higgs boson pair production at photon-photon colliders in Minimal Supersymmetric Standard Model has been extensively studied; however, there is a few works in THDM. The cross sections for the fusion processes $\gamma \gamma \to S_iS_j$ ($S_i=h^0, H^0, A^0$) have been computed in~\cite{Arhrib2009} and presented that a wide region of the parameter space where the cross section is two orders of magnitude larger than the ones of SM. In the context of the two-Higgs-doublet model type III, the production of neutral Higgs boson pairs at photon-photon colliders has been also studied in~\cite{Hern2012}. They have pointed up that the relevant processes are very sensitive to a general form of the Higgs potential which affect the triple-quartic couplings in the scalar sector. In the present work, the full set of one-loop contributions for the direct production of the pseudoscalar higgs pairs in $\gamma\gamma$ collisions are investigated in the framework of THDM taking into account both theoretical restrictions and experimental constraints from recent LHC data and other experimental results. The effect of individual contributions from each type of one-loop diagrams on the total cross-section are also examined in detail. Note that the results of the present study are consistent with those obtained in previous works.

The remainder of the paper is organized as follows. In Section \ref{sec:THDM}, a brief review is given for the THDM.
Section~\ref{sec:parameter} presents the experimental and theoretical constraints on parameter space of the THDM and four benchmark points scenarios which are consistent with these constraints.
In Section \ref{sec:cros}, the corresponding one-loop Feynman diagrams are presented and analytical expressions for the production cross section are briefly reviewed. The numerical evaluation method is then explained. In Section~\ref{sec:results},  numerical results are presented and the corresponding model parameter dependencies of the cross section are discussed in detail. Finally, in Section~\ref{sec:conc} the concluding remarks of the study are given.

\section{Review of the two Higgs doublet model}\label{sec:THDM}
For completeness, we first give a brief summary of the CP-conserving THDM here, including only those details relevant to this study. THDM has been extensively studied in the literature. The interested reader can refer to the reference~\cite{THDM2} as a comprehensive review of these models.

The THDM is the most minimal extension of the SM containing extra Higgs doublet fields. In the THDM,  the most general scalar potential being invariant under the SM electroweak gauge group ${\rm SU(2)}_L\otimes  {\rm  U(1)}_Y$, can be written as
\begin{equation} \label{eq:potential}
\begin{split}
V_{\text{THDM}}  = & m_1^2 |\Phi_1|^2 +m_2^2 |\Phi_2|^2
-\bigg[m_{12}^2 (\Phi_1^\dag \Phi_2) \\ & +{\rm h.c.}\bigg]
+\frac{\lambda_1}{2}(\Phi_1^\dag \Phi_1)^2+\frac{\lambda_2}{2}(\Phi_2^\dag \Phi_2)^2\\
& +\lambda_3(\Phi_1^\dag \Phi_1)(\Phi_2^\dag \Phi_2)
 +\lambda_4(\Phi_1^\dag \Phi_2)(\Phi_2^\dag \Phi_1) \\
&+ \bigg[\frac{\lambda_5}{2}(\Phi_1^\dag \Phi_2)^2
+\lambda_6(\Phi_1^\dag \Phi_1)(\Phi_1^\dag \Phi_2)\\
&+\lambda_7(\Phi_1^\dag \Phi_2)(\Phi_2^\dag \Phi_2)+{\rm h.c.}\bigg]
\end{split}
\end{equation}
where $\Phi_{1,2}$ are two complex scalar Higgs doublets and  $\lambda_i$ (i=1,...,7) are dimensionless quartic coupling parameters\footnote{In case of CP conservation, which allows the SM-like Higgs to be a CP-even scalar, the parameters in~\eqref{eq:potential} are required to be real.}. To respect some low energy observables
, the discrete $Z_2$ symmetry proposed by the Paschos-Glashow-Weinberg theorem~\cite{Glashow} is imposed to avoid tree-level flavor changing neutral currents. As a result, the $Z_2$ symmetry requires that $\lambda_{6,7}$ and $m_{12}^2$ must be zero. However, letting $m_{12}^2$ be non-zero, this symmetry can be softly broken. The charges under this symmetry are assigned to ensure that each type of fermion couples to only a single Higgs doublet. There are 4 types of THDMs, which are commonly called as type-I, type-II, type-III and type-IV of THDM, depending on the $Z_2$ assignment~\cite{THDM1,THDM2}. The way in which each Higgs doublet ($\Phi_{1,2}$) couples to the fermions in the allowed types which naturally conserve flavor is given in Table~\ref{tab:coupling}.
\begin{table}[h]
\caption{Couplings of $u$-type quarks, $d$-type quarks and charged leptons to Higgs doublets $\Phi_{1,2}$ in types allowed by the $Z_2$ symmetry. The subscript $i$ is a generation index.
\label{tab:coupling}}
\begin{ruledtabular}
  \begin{tabular}{rlll}
type& $u_i$ & $d_i$ & $\ell_i$ \\
\hline
  I & $\Phi_2$ & $\Phi_2$& $\Phi_2$ \\
 II & $\Phi_2$ & $\Phi_1$& $\Phi_1$ \\
III & $\Phi_2$ & $\Phi_2$& $\Phi_1$  \\
 IV & $\Phi_2$ & $\Phi_1$& $\Phi_2$ \\
\end{tabular}
\end{ruledtabular}
\end{table}
The types ``III'' and ``IV'' are also known as ``lepton-specific'' and ``flipped'', respectively.
In this study the numerical analysis will be carried out in the framework of the Type-I of THDM, in which only the doublet $\Phi_2$ interacts with both quarks and leptons like in SM.

After electroweak symmetry breaking, each scalar doublet acquires a vacuum expectation value $v_j$ such that $v=\sqrt{v_1^2+v_2^2}\approx 246~$GeV and
\begin{eqnarray}
\Phi_j  =  \left(\begin{array}{c}
  \phi_j^+ \\  \frac{1}{\sqrt{2}}(v_j+\rho_j+i \eta_j) \end{array}\right), (j=1,2),
\end{eqnarray}
where $\rho_j$ and $\eta_j$ are real scalar fields. 
The two Higgs doublets have initially 8 degrees of freedom. The three of them (Goldstone bosons $G^0$, $G^{\pm}$) are absorbed by the longitudinal components of the electroweak gauge bosons $Z$ and $W^\pm$. The remaining five are physical Higgs fields, a CP-odd pseudoscalar
$A^0$, two CP-even $h^0$ and $H^0$,  and two charged scalars $H^+$ and $H^-$. The relevant mass eigenstates are determined by orthogonal transformations, in which the angles $\alpha$ and $\beta$ govern the mixing between mass eigenstates in the CP-even sector and CP-odd/charged sectors, respectively.

For any given value of $\tan\beta$, $m_1^2$ and $m_2^2$ are calculated by the minimization conditions of potential in a minimum of the vacuum. The mass parameters $m^2_{1,2}$ and quartic couplings $\lambda_1$--$\lambda_5$ can be expressed in terms of the physical masses $m_h$, $m_H$, $m_A$, $m_{H^\pm}$, along with the ratio of vacuum expectation values, namely $\tan\beta=v_2/v_1$, and the neutral sector mixing term $\sin(\beta-\alpha)$. The soft $Z_2$ symmetry breaking parameter $m^2_{12}$ can be written as
\begin{equation}\label{eq:m122}
m^2_{12}=\frac{1}{2} \lambda_5 v^2 \sin \beta \cos\beta = \frac{\lambda_5}{2\sqrt{2} G_F} \frac{\tan\beta}{1+\tan^2\beta},
\end{equation}
where the second equality is valid at the tree level. Setting $\lambda_{6}$ and $\lambda_{7}$ to zero to respect the discrete $Z_2$ symmetry and working in the “physical basis”, $m_{12}^2$,  $\tan\beta$, mixing angle $\alpha$ and four physical masses of the Higgs bosons can be determined to specify the model completely. Consequently, in the Higgs sector of the THDM, there are seven independent parameters. From the above potential, Equation~(\ref{eq:potential}), the triple and quartic scalar couplings can be derived as a function of the masses of neutral and charged Higgs
$m_{h^0}$, $m_{H^0}$, $m_{A^0}$, $m_{H^\pm}$, and $\tan\beta$, $\alpha$
and $m_{12}^2$ as follows\footnote{The short-hand notation $c_x$ and $s_x$ are used for
$\cos(x)$ and $\sin(x)$, respectively. For example, $c_{\al+\be}=cos(\al+\be)$ for $x=\al+\be$.}:
\begin{equation}\label{eq:lambda}
\begin{split}
\lambda_{h^0h^0h^0}^{THDM}=& \frac{-3g}{2m_W s^2_{2\be}}\bigg[(2c_{\al+\be} + s_{2\al}
  s_{\be-\al})s_{2\be} m^2_{h^0} \\
  &- 4(c^2_{\be-\al} c_{\be + \al}) m^2_{12}\bigg],
\end{split}
\end{equation}
\begin{equation}\label{eq:lambda2}
\begin{split}
\lambda_{H^0h^0h^0}^{THDM} =& -\frac{g c_{\be-\al}}{2 m_W s^2_{2\be}}\bigg[
  (2 m^2_{h^0} + m^2_{H^0}) s_{2\al} s_{2\be} \\
  &-2 (3 s_{2\al}-s_{2\be})
  m^2_{12}\bigg],
\end{split}
\end{equation}
\begin{equation}\label{eq:lambda3}
\begin{split}
\lambda_{h^0H^0H^0}^{THDM} =& \frac{g s_{\be-\al}}{2 m_W s^2_{2\be}}\bigg[
  (m^2_{h^0} + 2 m^2_{H^0}) s_{2\al} s_{2\be} \\
  &+ 2 (3 s_{2\al}+s_{2\be})
  m^2_{12}\bigg],
\end{split}
\end{equation}
\begin{equation}\label{eq:lambda4}
\begin{split}
\lambda_{h^0A^0A^0}^{THDM} =& \frac{g}{2m_W}\bigg[
  (m^2_{h^0} -2 m^2_{A^0} )s_{\be-\al} \\
  &- \frac{2c_{\be + \al}}{s^2_{2\be}}(m^2_{h^0} s_{2\be}-
  2 m^2_{12})\bigg],
\end{split}
\end{equation}
\begin{equation}\label{eq:lambda5}
\begin{split}
\lambda_{ h^0 H^\pm H^\mp}^{THDM} =& \frac{g}{2 m_W}\bigg[
  (m^2_{h^0} - 2 m^2_{H^\pm})s_{\be-\al} \\
  &-\frac{2c_{\be + \al}}{s^2_{2\be}}(m^2_{h^0} s_{2\be}-
 2 m^2_{12})\bigg],
 \end{split}
\end{equation}
\begin{equation}\label{eq:lambda6}
\begin{split}
\lambda_{A^0G^0h^0}^{THDM} = &\frac{g c_{\be-\al}}{2m_W}\bigg[ m^2_{A^0}
- m^2_{h^0}\bigg],\quad\\ \lambda_{A^0H^-G^+}^{THDM} =& \frac{g }{2m_W}\bigg[ m^2_{A^0}
- m^2_{H^-}\bigg],
\end{split}
\end{equation}
\begin{equation}\label{eq:lambda7}
\begin{split}
\lambda^{THDM}_{A^0 A^0 H^- H^+}& =\frac{ -g^2 }{4
m_{W}^2 s^2_{2\beta}} \bigg[ m^2_{H^0}
(c_{\beta-\alpha}s_{2\beta}-2s_{\beta+\alpha})^2 \\
+ m^2_{h^0}&
(2c_{\beta+\alpha} - s_{2\beta}s_{\beta-\alpha})^2+8 \frac{m^2_{12} c^2_{2\beta}}{s_{2\beta}}\bigg]
\end{split}
\end{equation}
where the parameter $g=e/sin\theta_W$ is the $SU(2)$ gauge coupling constant and $m_W$ is the mass of $W$ boson. These triple Higgs couplings are independent of the Yukawa types used, because they follow from the scalar THDM potential. All these couplings have a strong dependence on the mixing angles $\alpha$ and $\beta$, the physical higgs masses, and the soft breking term $m_{12}^2$ parameter. In this study, in particular, triple Higgs couplings and couplings of the (pseudo)scalar to gauge bosons are interested. The (pseudo)scalar-gauge couplings
\begin{equation}
\begin{split}
&\lambda_{ h^0 W^\pm W^\mp}=g m_W s_{\beta-\alpha}, \lambda_{h^0G^\pm W^\mp  }=\frac{g}{2} s_{\beta-\alpha} , \\
& \lambda_{ H^0 W^\pm W^\mp}=g m_W c_{\beta-\alpha}, \lambda_{H^0G^\pm W^\mp  }=\frac{g}{2} c_{\beta-\alpha}
\end{split}
\end{equation}
are proportional to $\cos(\beta-\alpha)$ or $\sin(\beta-\alpha)$, while
$\lambda_{A^0 H^\pm W^\mp}=e/2s_W$
is independent of the THDM angles. Contrary to the CP-even Higgs bosons $h^0$ and $H^0$, pseudoscalar Higgs boson, due to its CP-odd nature, does not couple to pairs of $ZZ$ and $W^+W^-$. Therefore, Z-boson and W-boson loop diagrams do not contribute to pseudoscalar Higgs boson production at one-loop level.

\begin{table*}[]
\caption{Selected BPs using Higgs data for 2HDM type-I with alignment limit. For all BPs CP-even Higgs mass is fixed as $m_h=125.18$ GeV and tan$\beta $ is set to 10. All BPs are still allowed by the searches for additional Higgs bosons at the LHC.}\label{BP}
\begin{ruledtabular}
\begin{tabular}{cccccccc}
 BPs&$m_{h^0}$ (GeV)&$m_{A^0}$ (GeV)&$m_{H^0}$  (GeV)&$m_{H^\pm}$ (GeV)&$m_{12}^2$ (GeV$^2$)&tan$\beta $ &sin($\beta-\alpha$)  \\
\hline
\textbf{BP1}   &\multirow{4}{*}{125.18} &150  &150 &150 &2000 &\multirow{4}{*}{10}&\multirow{4}{*}{1}\\
\textbf{BP2}   &                        &200  &150 &250 &2000 &                   & \\
\textbf{BP3}   &                        &250  &150 &250 &2000 &                   & \\
\textbf{BP4}   &                        &250  &250 &300 &6000 &                   & \\
\end{tabular}
\end{ruledtabular}
\end{table*}

In THDM, a decoupling limit appears when $\cos(\beta-\alpha)=0$ and $m_{H^0,A^0,H^\pm}\gg m_Z$ \cite{Gunion}. In this limit, the coupling of the higgs boson $h^0$ to SM
particles entirely look alike the SM Higgs couplings which include the coupling $h^0h^0h^0$.
Furthermore, there also is an alignment limit~\cite{Carena}, where the CP-even Higgs boson
$h^0$ ($H^0$) looks like SM Higgs boson if $\sin(\beta-\alpha)\to 1$ ($\cos(\beta-\alpha)\to 1$). In the decoupling or alignment limit with $\alpha = \beta-\pi/2$, the some triple Higgs couplings turn into the following form
\begin{equation}
\begin{split}
&  \lambda_{h^0h^0h^0}^{THDM} = \frac{-3g}{2m_W}m^2_{h^0} = \lambda_{hhh}^{SM},~~\lambda_{H^0h^0h^0}^{THDM} =0,\\
& \lambda_{h^0H^0H^0}^{THDM} = \frac{g}{m_W}\bigg[
  \bigg(\frac{2m^2_{12}}{s_{2\be}}-m^2_{H^0}\bigg) - \frac{m^2_{h^0}}{2}
  \bigg],\\
&~~\lambda_{h^0A^0A^0}^{THDM}  = \frac{g}{m_W}\bigg[
  \bigg(\frac{2m^2_{12}}{s_{2\be}}-m^2_{A^0}\bigg)-\frac{m^2_{h^0}}{2}\bigg],\\
&  \lambda_{h^0H^\pm H^\mp}^{THDM} =  \frac{g}{m_W}\bigg[
  \bigg(\frac{2m^2_{12}}{s_{2\be}}-m^2_{H^\pm}\bigg)-\frac{m^2_{h^0}}{2}\bigg].
\end{split}
\end{equation}

\section{Parameter Setting and Constraints on THDM}\label{sec:parameter}
The parameter space of the scalar THDM potential is reduced both by the results of experimental searches as well as by theoretical constraints. The THDM are subjected to several theoretical constraints such as potential stability, perturbativity and unitarity. For ensuring vacuum stability of the THDM, the ${V}_{\rm THDM}$ must be bounded from below. In other words, ${V}_{\rm THDM}\geq 0$ must be maintained for all directions of $\Phi_1$ and $\Phi_2$. This constraint puts the following conditions on the parameters $\lambda_i$ \cite{BFB1,BFB2}:
\begin{equation}
\begin{split}
&\lambda_1>0, \lambda_2>0, \lambda_3 + 2
 \sqrt{\lambda_1 \lambda_2} > 0,\\
&\lambda_3 + \lambda_4 - |\lambda_5| > 2 \sqrt{\lambda_1 \lambda_2}.
\end{split}
\end{equation}
There is also another set of constraints which imposes that the perturbative unitarity must be satisfied for scattering of longitudinally polarized gauge bosons and Higgs bosons. They can be found in~\cite{unitarity1,unitarity2}. Furthermore, the scalar potential must be perturbative by imposing that all quartic coefficients satisfy $|\lambda_{1,2,3,4,5}| \leq 8 \pi$.

Besides the above theoretical constraints, the THDM has the current constraints resulting from direct observations at the LHC and indirect experimental limits from $B$~physics observables. In the Type-I of THDM, pseudoscalar Higgs mass regions such as $310<m_A<410$ GeV for $m_H = 150$ GeV, $335<m_A<400$ GeV for $m_H = 200$ GeV, $350<m_A<400$ GeV for $m_H = 250$ GeV at $tan\beta = 10$ has been excluded by the LHC experiment~\cite{ATLAS2}. Moreover, the limit $m_A > 350$ is put on the pseudoscalar Higgs mass for $tan\beta < 5$~\cite{ATLAS3} and the mass range $170<m_H<360$ GeV is excluded for $tan\beta < 1.5$ in the Type-I~\cite{ATLAS4}.

In type-II and IV of THDM, the data from the measurement of the branching ratio $b\to s\gamma$ puts constraints on the charged Higgs mass $m_{H^\pm}>580$~GeV~\cite{Misiak2017,Misiak2015} for $tan\beta \geq 1$. However, for the other types of THDM, this bound is much lower~\cite{Enomoto:2015wbn}. In type-I and III of THDM, as long as $tan\beta\geq 2$, the charged Higgs bosons are possible to be as light as 100 GeV \cite{Enomoto:2015wbn,Arhrib:2016wpw} while being compatible with LHC and LEP bounds as well as with all $B$ physics restrictions \cite{Aad:2014kga,Khachatryan:2015qxa,Khachatryan:2015uua,
Aad:2013hla, Abbiendi:2013hk,Akeroyd:2016ymd}. Moreover, there is no exclusion around $sin(\beta-\alpha)= 1$ for $m_{A,H,H^\pm} = 500$ GeV in the Type-I THDM according to a review of LEP, LHC and Tevatron results~\cite{Moretti}.

In order to illustrate the effect of the new physics, we have chosen four benchmark points (BPs) scenarios which are consistent with theoretical and experimental constraints as shown in Table~\ref{BP}. The benchmark points are constructed on type I of THDM with an alignment limit $sin(\beta-\alpha)\to 1$, and hence the CP-even higgs $h^0$ is a SM-like Higgs. Its mass is fixed as $m_h=125.18$ GeV~\cite{PDG}. The value of tan$\beta $ is set to 10 for all benchmark scenarios which result in a remarkable enhancement in the assumed scalar Higgs boson decay channel.
The potential stability, perturbativity and unitarity of each bencmhark point have been checked with the help of {\tt 2HDMC 1.7.0}~\cite{2HDMC,2HDMC2}. The oblique parameters $S, T$ and $U$ are calculated with {\tt 2HDMC 1.7.0} and are required to fall within the $95\%$ CL ellipsoid based on 2018 PDG values~\cite{PDG}. The considered benchmark scenarios are also consistent with the limits obtained from various searches for additional Higgs bosons at the LHC, and by the requirement that the CP-even higgs $h^0$ should match the properties of the observed Higgs-like boson. The constraints are checked by the public codes {\tt  HiggsBounds 4.3.1}~\cite{HBounds} and {\tt HiggsSignals 1.4.0}~\cite{HSignals} with results of 86 analyses. In Table~\ref{tab:branching}, the dominant branching ratios of CP-odd Higgs $A^0$, which are computed by using {\tt 2HDMC 1.7.0}, are listed for selected BPs.
\begin{table}[!tbph]
\medskip
\centering\renewcommand{\arraystretch}{1.0}
\caption{The dominant branching ratios (BRs) of CP-odd Higgs $A^0$ for selected BPs, where BRs values which are less than $10^{-4}$ are not shown.}\label{tab:branching}
\begin{tabular}{|l|c|c|c|c|}
\hline
 \textbf{BR}&\textbf{BP1} & \textbf{BP2} & \textbf{BP3}  &\textbf{BP4}  \\
\hline\hline
$A^0 \rightarrow gg$            &3.06$\times10^{-1}$  &1.67$\times10^{-1}$   &0.09$\times10^{-2}$  &\cellcolor{LightCyan} 5.93$\times10^{-1}$  \\
$A^0 \rightarrow b\bar{b}$      &\cellcolor{LightCyan}6.03$\times10^{-1}$ &1.78$\times10^{-1}$   &0.05$\times10^{-2}$  &3.49$\times10^{-1}$\\
$A^0 \rightarrow c\bar{c}$      &2.78$\times10^{-2}$  &0.81$\times10^{-2}$   &$<10^{-4}$  &1.61$\times10^{-2}$  \\
$A^0 \rightarrow \tau^+ \tau^-$ &6.15$\times10^{-2}$  &1.92$\times10^{-2}$   &$<10^{-4}$  &3.94$\times10^{-2}$  \\
$A^0 \rightarrow Z^0 H^0$       &$-$                  &\cellcolor{LightCyan} 6.28$\times10^{-1}$   &\cellcolor{LightCyan} 9.98$\times10^{-1}$  &$-$ \\
\hline
\end{tabular}
\centering
\end{table}
The most dominant ratios are marked by turquoise blue: $A^0 \rightarrow b\bar{b}$ for BP1, $A^0 \rightarrow Z^0 H^0$ for BP2 and BP3, and $A^0 \rightarrow gg$ for BP4.

\section{Analytical expressions for the production cross section}\label{sec:cros}
\begin{figure*}[!tbph]
    \begin{center}
\includegraphics[width=0.9\linewidth]{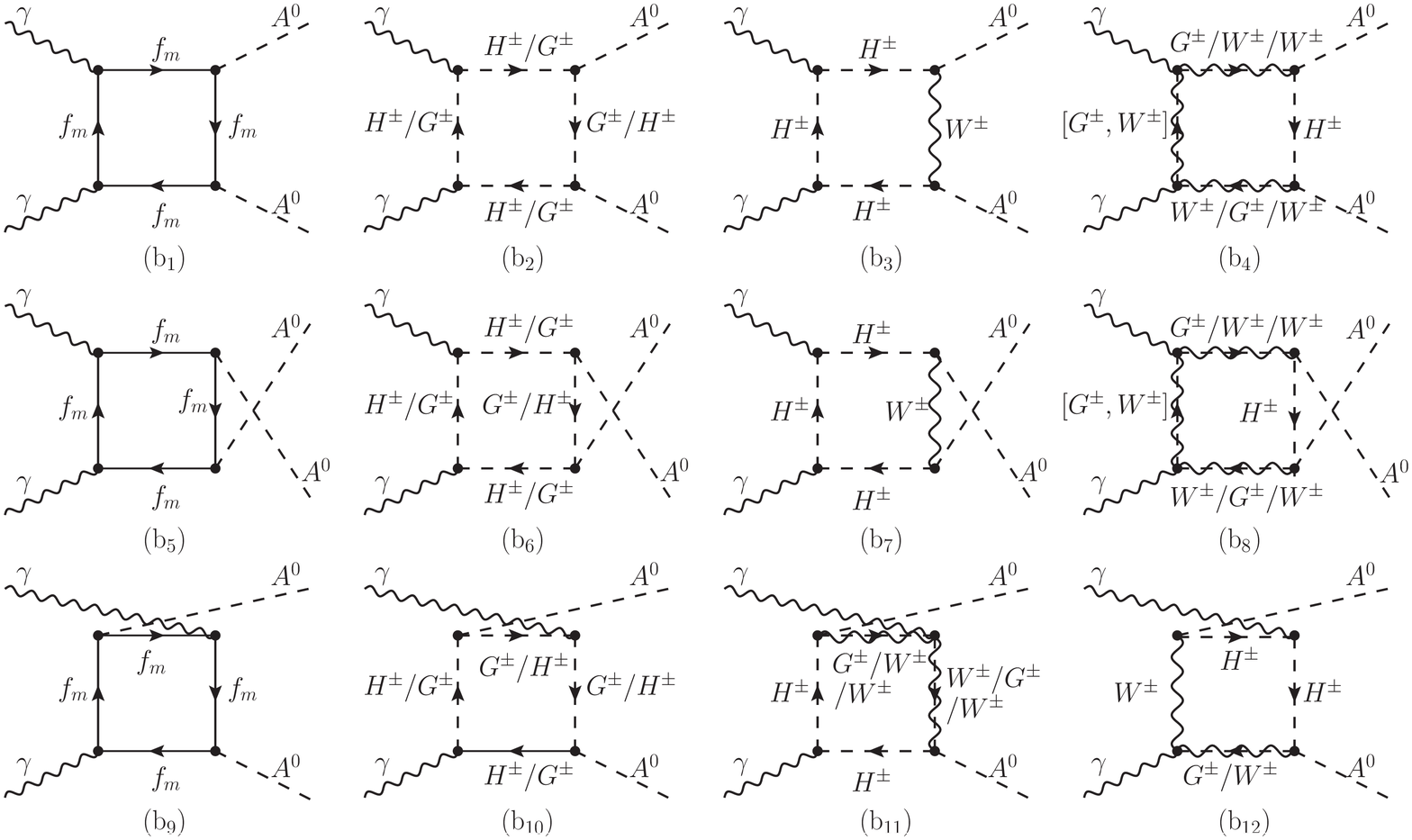}
     \end{center}
\caption{Box-type diagrams contributing to the process $\gamma\gamma\rightarrow A^{0}A^{0}$
at one-loop level. Here, the label ${f}_m$ represents to fermions of $e,\mu, \tau ,u, d, c, s, t$ and $b$. Dashed-lines in loops represent charged Higgs bosons $H^\pm$ and charged Goldstone boson $G^\pm$, and wavy-lines in loops represent W bosons.}\label{fig:fig1}
\end{figure*}
\begin{figure*}[!ht]
    \begin{center}
\includegraphics[width=0.9\linewidth]{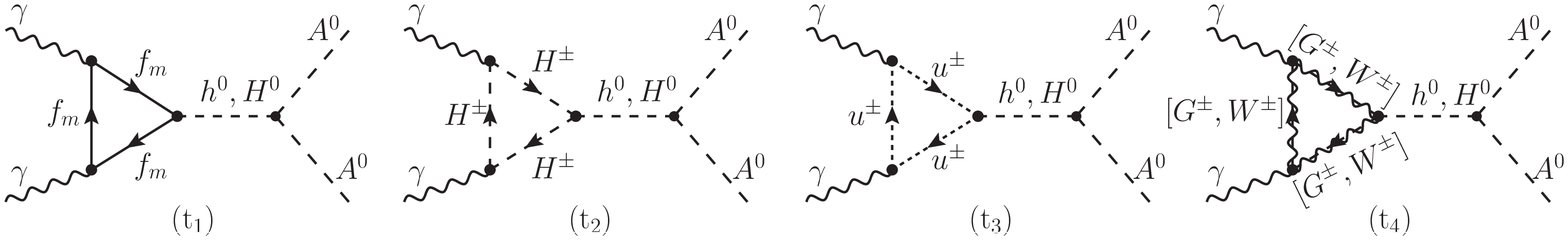}
     \end{center}
\caption{Triangle-type diagrams contributing to the process $\gamma\gamma\rightarrow A^{0}A^{0}$
at one-loop level. Here, the label ${f}_m$ refers to fermions of $e,\mu, \tau ,u, d, c, s, t$ and $b$. Dashed-lines in loops represent charged Higgs bosons $H^\pm$ and charged Goldstone boson $G^\pm$, and wavy-lines in loops represent W bosons.} \label{fig:fig2}
\end{figure*}
\begin{table*}[!bht]
\caption{Triple and quartic Higgs couplings and couplings of the (pseudo)scalar to W-boson which are included in each type of diagrams.}\label{tab:triplediagram}
\begin{tabular}{|c|l|c|c|c|c|}
\hline
&\textbf{Couplings}&\textbf{Box-type} & \textbf{Triangle-type} & \textbf{Bubble-type}  &\textbf{Quartic-type}  \\
\hline\hline
\parbox[t]{2mm}{\multirow{4}{*}{\rotatebox[origin=c]{90}{SSS}}} &$\lambda_{[h^0,H^0] A^0 A^0 }$   &           &\checkmark (t$_{1,2,3,4}$)&\checkmark (q$_{1,2,3}$)&       \\
&$\lambda_{H^+ H^-[h^0,H^0]}$   &           &\checkmark (t$_2$) &\checkmark (q$_3$)&       \\
&$\lambda_{G^+ G^-[h^0,H^0]}$   &           &\checkmark (t$_4$) &\checkmark (q$_3$)&       \\
&$\lambda_{H^+G^- A^0 }$          &\checkmark (b$_{2,4,6,8,{10},{11},{12}}$)&           &       &\checkmark (q$_{9,{10},{11},{12},{13}}$) \\
\hline
\multirow{3}{*}{\rotatebox[origin=c]{90}{\parbox[c]{1cm}{\centering WSS\\WWS}}} &$\lambda_{H^+W^- A^0 }$          &\checkmark (b$_{3,4,7,8,{11},{12}}$)&           &       &\checkmark (q$_{9,{10},{11},{12},{14}}$)       \\
&$\lambda_{W^+W^-[h^0,H^0]}$      &                      &\checkmark (t$_{4}$)  &\checkmark (q$_{3}$)       &       \\
&$\lambda_{G^+W^- [h^0,H^0]}$     &                      &\checkmark (t$_{4}$)             &       &       \\
\hline
\multirow{3}{*}{\rotatebox[origin=c]{90}{\parbox[c]{1cm}{\centering SSSS\\WWSS}}}&$\lambda_{H^+H^- A^0 A^0}$     &                      &             &\checkmark (q$_{4}$) &\checkmark (q$_{7}$)       \\
&$\lambda_{G^+G^- A^0 A^0}$     &                      &             &\checkmark (q$_{4}$)  &\checkmark (q$_{8}$)       \\
&$\lambda_{W^+W^- A^0 A^0}$     &                      &             &\checkmark (q$_{4}$) &\checkmark (q$_{8}$)       \\
\hline
\end{tabular}
\end{table*}
\begin{figure*}[!ht]
    \begin{center}
\includegraphics[width=0.9\linewidth]{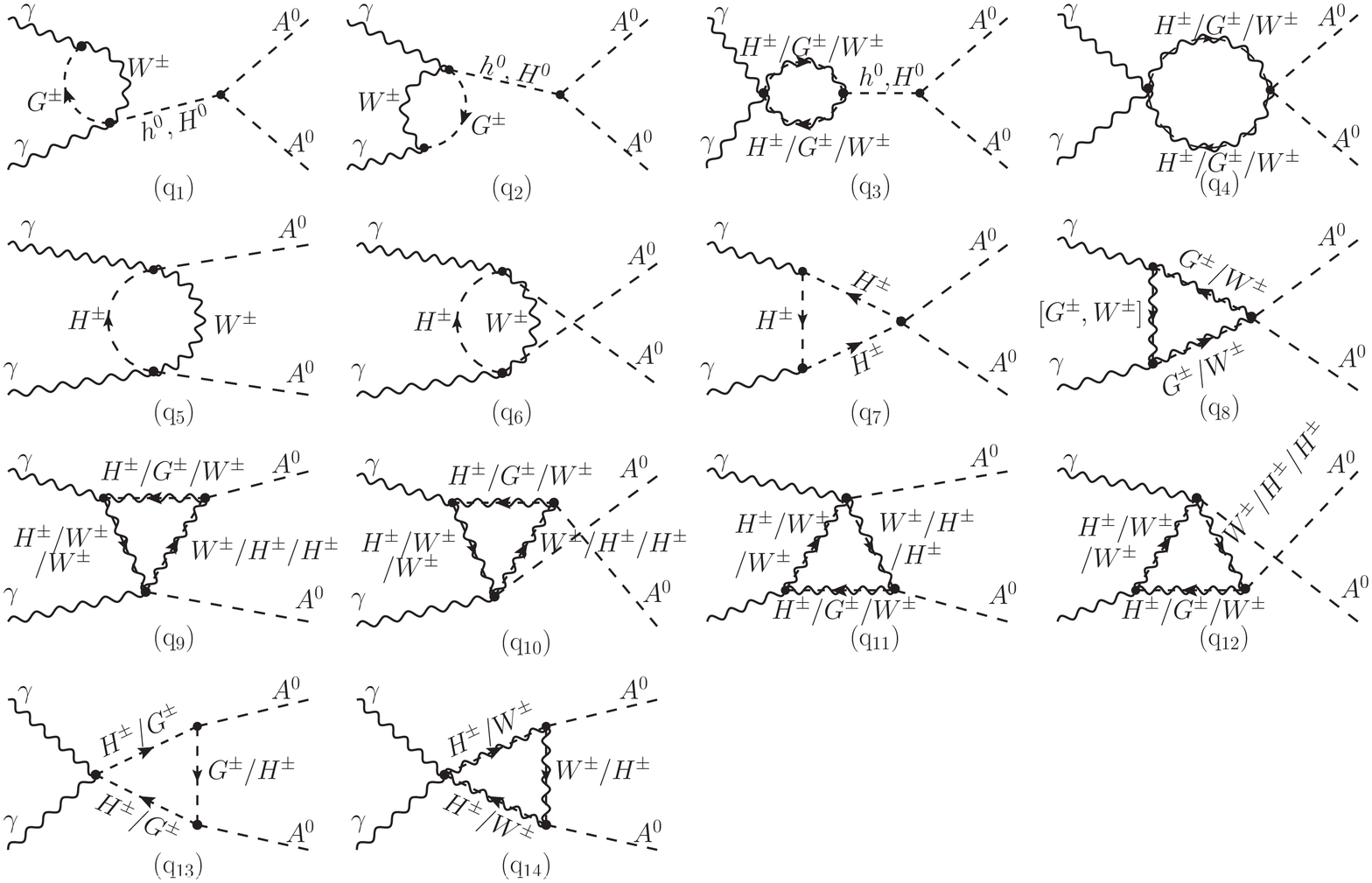}
     \end{center}
\caption{Quartic interaction diagrams contributing to the process $\gamma\gamma\rightarrow A^{0}A^{0}$
at one-loop level. Dashed-lines in loops represent charged Higgs bosons $H^\pm$ and charged Goldstone boson $G^\pm$, and wavy-lines in loops represent W bosons.}\label{fig:fig3}
\end{figure*}
The process of the pseudoscalar Higgs boson pair production in photon-photon collision is denoted by
\begin{equation} \label{eq:gammagammaAA}
\gamma(p_1)\gamma(p_2)\rightarrow A^{0}(k_1) A^{0}(k_2),
\end{equation}
where after each particle, as usual, its 4-momenta is written in parentheses. This subprocess has no an amplitude at tree-level, and has one-loop level amplitude in the lowest order. A full set of one-loop level Feynman diagrams\footnote{Note that Feynman diagrams have been plotted by using~\texttt{JaxoDraw}~\cite{JaxoDraw}.} contributing to the process~\ref{eq:gammagammaAA} in the THDM is generated by the \texttt{FeynArts} \cite{Feynarts}. They are shown in Figs.~\ref{fig:fig1} to~\ref{fig:fig3}, and also the process has another set of Feynman diagrams that are not given in the figures where particles in loops are flowing counterclockwise.
The square bracket $[G,W]$ means that the loop contains all possible combinations of the particles $G$ and $W$.

Any one-loop amplitude can be written as a linear sum of bubble, box, triangle, and tadpole one-loop integrals. According to the loop-correction type, the diagrams of $\gamma\gamma\rightarrow A^{0}A^{0}$ at one-loop level can be classified into three kinds of groups, which are called as the triangle-type, the box-type, and the quartic coupling-type diagrams. Figure~\ref{fig:fig1} shows all possible box-type diagrams, which have the loops of charged leptons and quarks of three generations, bosons of $G^\pm$, $W^\pm$, and $H^\pm$. These are $t$- and $u$-channel diagrams. Figure~\ref{fig:fig2} shows all triangle-type diagrams which consist of triangle vertices (t$_{1-4}$) attached to the final-state via an intermediate Higgs bosons $h^0$ or $H^0$. Finally, Figure~\ref{fig:fig3} shows all possible quartic coupling-type diagrams which include bubbles (q$_{1-3}$) attached to the final-state via an intermediate Higgs bosons $h^0$ or $H^0$, bubbles loop (q$_{4-6}$) and triangle loop (q$_{7}$, q$_{14}$) of the bosons $G^\pm$,  $H^\pm$, and $W^\pm$ directly attached to the final-state. The diagrams t$_{1-4}$ and q$_{1-3}$ are s-channel diagrams. The resonant effects appear only triangle diagrams (t$_{1-4}$) and in the bubbles-type (q$_{1-3}$) due to the intermediate neutral Higgs bosons.

Note that the Feynman diagrams of the process $\gamma\gamma\rightarrow H^{0}H^{0}$ and $\gamma\gamma\rightarrow H^{0}A^{0}$ are almost the same as those of the $\gamma\gamma\rightarrow A^{0}A^{0}$ considered in this study. Therefore, any result to be obtained for the process $\gamma\gamma\rightarrow A^{0}A^{0}$ can be also applied to these processes, depending parameters of model.

Table~\ref{tab:triplediagram} shows the triple and quartic Higgs couplings and couplings of the (pseudo)scalar to W-boson which are included in each type of diagrams. Feynman diagrams are dominated by triple couplings $\lambda_{H^+G^- A^0}$ and  $\lambda_{H^+W^- A^0 }$ which are independent of the THDM angles. Owing to the CP nature of $A^0$, the box-type diagrams are rather sensitive to the coupling $\lambda_{H^+G^- A^0}$ which does not have neither a $\tan\beta$ nor a $m_{12}^2$ dependence. Triple couplings $\lambda_{[h^0,H^0] A^0 A^0 }$, $\lambda_{H^+ H^-[h^0,H^0]}$ and $\lambda_{G^+ G^-[h^0,H^0]}$ only appear at s-channel diagrams and they are amplified by resonance effects due to neutral Higgs bosons. Diagrams $q_4$, $q_7$ and $q_8$ in Figure~\ref{fig:fig3} are sensitive to quartic couplings $\lambda_{H^+H^- A^0 A^0}$ and $\lambda_{G^+G^- A^0 A^0}$ which are proportional to mixing angles and the mass parameter $m_{12}^2$.

The amplitude of the process $\gamma\gamma\to A^0 A^0$ at one-loop level can be computed by summing all unrenormalized reducible and irreducible contributions. Consequently, one can obtain finite and gauge invariant results. Therefore, the renormalization for ultraviolet divergence does not need to be taken into account. The corresponding matrix element\footnote{In this study, an expression of matrix element is not explicitly presented because it is too lengthy to include here.} is calculated as a sum over triangle-type, box-type, bubble-type, and quartic-type contributions:
\begin{equation}\label{eq:totalM}
{\cal M}= {\cal M}_{box}+ {\cal M}_{quartic}+{\cal M}_{bubble}+{\cal M}_{triangle},\
\end{equation}
where a relative sign $(-1)^{\delta_{ij}}$ are written between the amplitudes of one diagram and its counterpart emerging by interchanging the final states. The total cross section of $\gamma\gamma\rightarrow A^{0}A^{0}$ are given by
\begin{equation} \label{eq:totalsigma}
\hat{\sigma}(\hat{s}_{\gamma\gamma},\gamma\gamma\rightarrow A^{0}A^{0})=\frac
{1}{32\pi \hat{s}^{2}_{\gamma\gamma}}\int_{\hat
t^{-}}^{\hat t^{+}}d\hat t \overline{\sum} |{\cal M}|^2,
\end{equation}
where the bar over the sum refers to the average over initial spins, and
$\hat{t}^\pm=(m_A^2-\hat{s}_{\gamma\gamma}/2) \pm\bigl(\sqrt{(\hat{s}_{\gamma\gamma}-2m_A^2)^2-4m_A^4}\bigr)/2$.

The $\gamma\gamma $ collision can be performed at the facility of the next generation of TeV-class linear colliders such as the ILC and the CLIC. Then, the $\gamma\gamma\to A^0 A^0$ is produced as a subprocess of $e^-e^+$ collisions at the linear colliders. The total cross section of the process $ e^+e^-\rightarrow\gamma\gamma\rightarrow A^{0}A^{0}$, could be obtained by folding the cross section of the $\gamma\gamma\to A^0 A^0$ with the photon luminosity
\begin{equation}
\frac{dL_{\gamma\gamma}}{dz}=2z\int_{z^2/x_{max}}^{x_{max}}\frac{dx}{x}F_{\gamma/e}(x)F_{\gamma/e}\left(\frac{z^2}{x}\right)\,,
\end{equation}
as follows
\begin{align} \label{eq:total_cross}
\begin{split}
&\sigma(s, e^+e^-\rightarrow\gamma\gamma\rightarrow  A^{0}A^{0})=\\
&\int_{(2m_{A^0})/\sqrt{s}}^{x_{max}} dz \frac{dL_{\gamma\gamma}}{dz}~ \hat{\sigma}( \gamma\gamma\rightarrow  A^{0}A^{0};\; \hat{s}_{\gamma\gamma}=z^2s ).
\end{split}
\end{align}
where $F_{\gamma/e}(x)$ is the photon structure function. The photon spectrum is qualitatively better for larger values of fraction $x$ of the longitudinal momentum of the electron beam. However, for $x > 2(1+\sqrt{2})\approx 4.8$, the high-energy photons can disappear through the pair production of $e^+e^-$ in its collision with a following laser photon. The energy spectrum of the photon supplied as Compton backscattered photon off the electron beam~\cite{Telnov} is used for photon structure function of this study.

The numerical evaluation for both $\gamma\gamma\rightarrow A^{0}A^{0}$ and  $e^+e^-\rightarrow\gamma\gamma\rightarrow A^{0}A^{0}$ is carried out by the help of the Mathematica packages as follows: The relevant amplitudes are generated by \texttt{FeynArts}~\cite{Feynarts},  the analytical expressions of the squared matrix elements are provided by \texttt{FormCalc} \cite{Hahn}, and the necessary one-loop scalar integrals are evaluated by \texttt{LoopTools} \cite{loop}. The integration over phase space of $2 \rightarrow 2$ is numerically evaluated by using CUBA library.
For the photon structure function, Compton backscattered photons which are interfaced by the \texttt{CompAZ} code \cite{Compaz} are used. Using the methods described above, we have previously studied the production of neutralino pairs in the photon-photon collision and found significant results~\cite{Demirci}.

\section{Numerical Results And Discussion} \label{sec:results}
In this section, the numerical predictions for the direct pair production of the pseudoscalar Higgs boson at a photon-photon collision are presented in detail, taking into account a full set of one-loop level Feynman diagrams. During our calculations, the cancellation of divergences appearing in the loop contributions have been numerically checked, giving, finite results without the need of the renormalization procedure. The integrated cross section $\hat{\sigma}(\gamma\gamma\rightarrow A^{0}A^{0})$ is analyzed as a function of the center-of-mass energy $\sqrt{\hat{s}}_{\gamma\gamma}$, focusing on the individual contributions from each type of diagram and on the polarization configurations of the incoming photons, for representative BPs given in Table~\ref{BP}. The dependencies of $\hat{\sigma}(\gamma\gamma\rightarrow A^{0}A^{0})$ on the plane of $m_{12}^2-\tan \beta$ and $m_A-\tan \beta$ are also investigated. Furthermore, the total cross section $\sigma(e^+e^-\rightarrow\gamma\gamma\rightarrow A^{0}A^{0})$ is numerically evaluated as a function of the center-of-mass energy (for BPs given in Table~\ref{BP}) and as a function of $m_A$ for several Higgs mass hierarchy.

\begin{figure*}[!ht]
    \begin{center}
\includegraphics[scale=0.38]{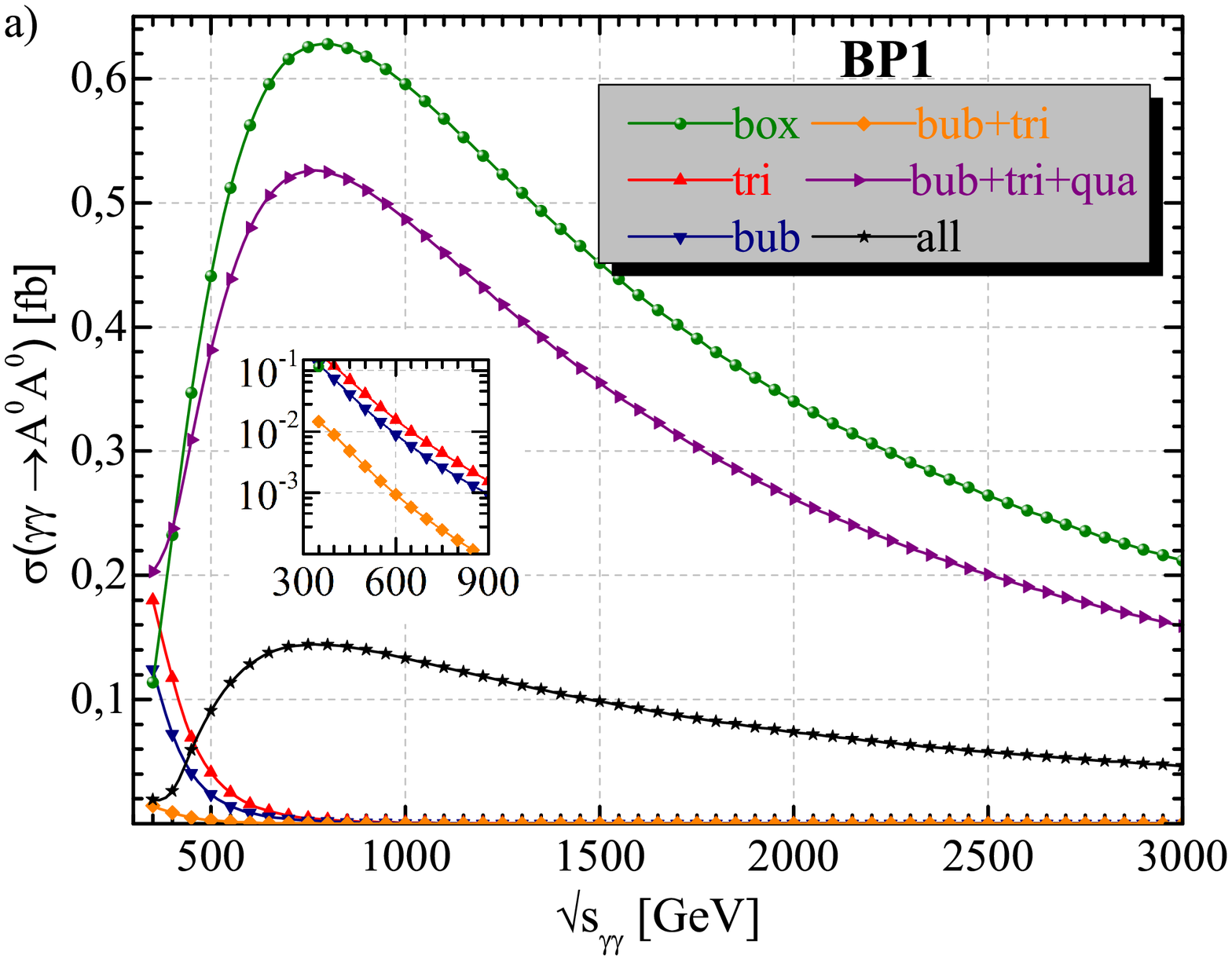}
\includegraphics[scale=0.38]{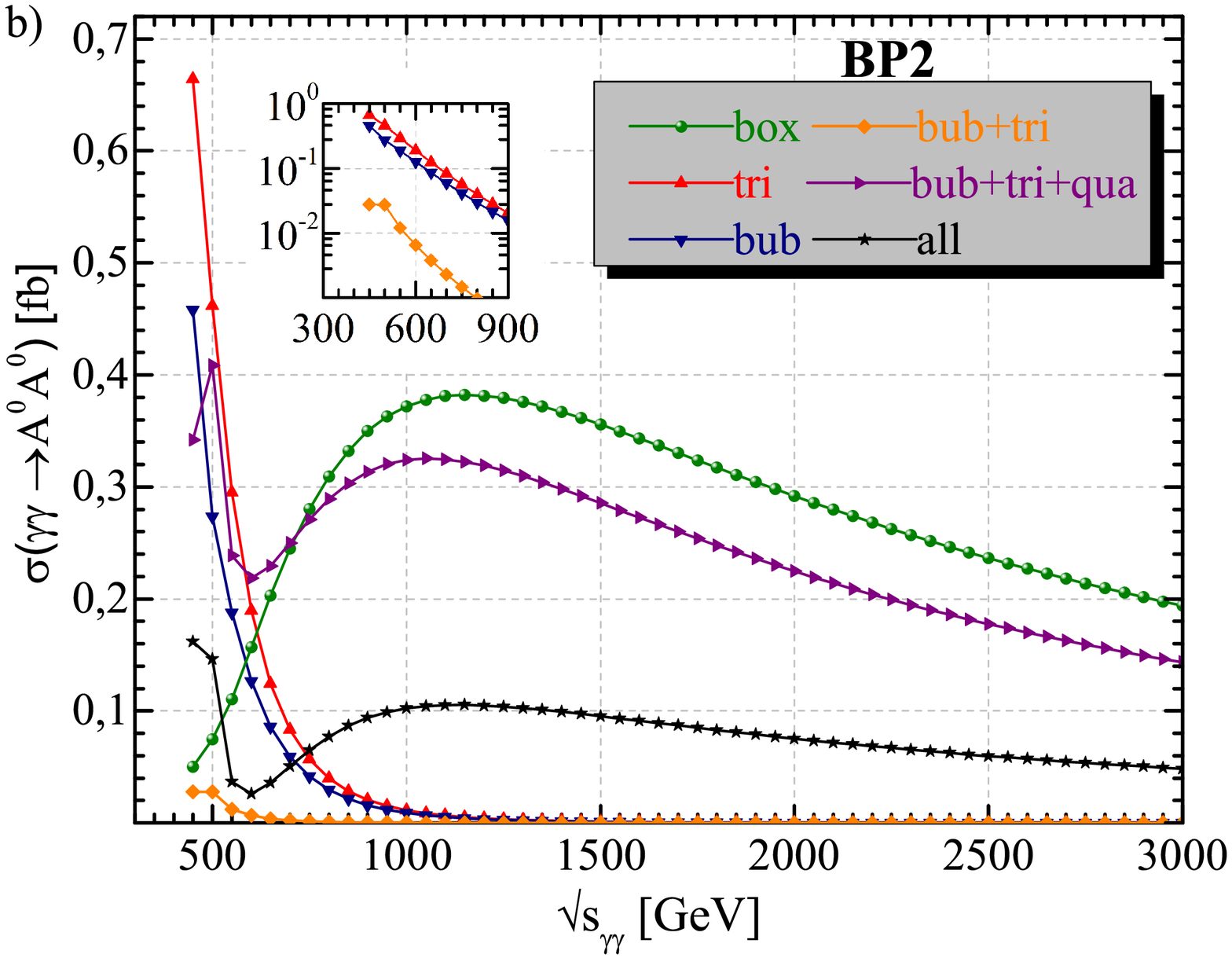}
\includegraphics[scale=0.38]{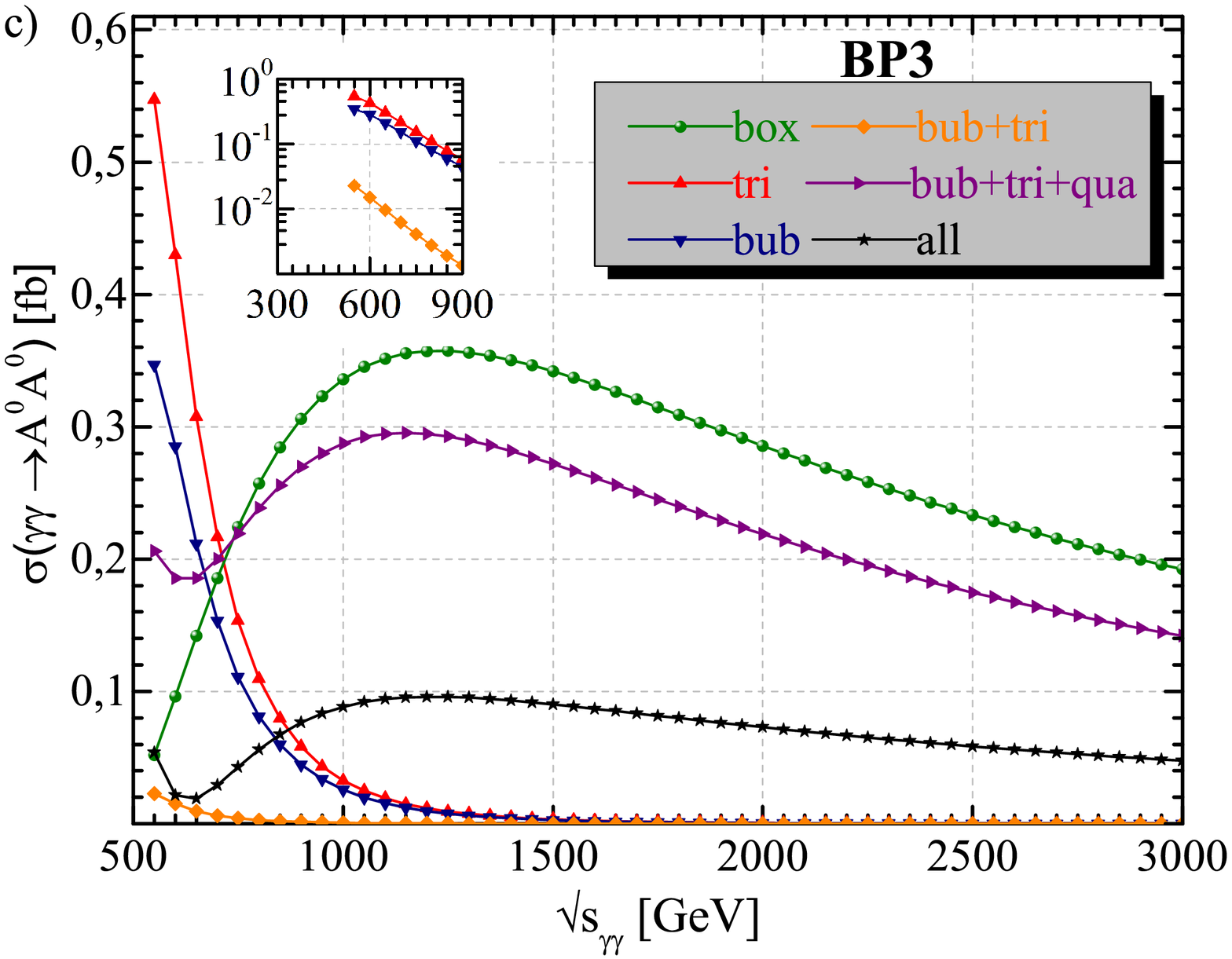}
\includegraphics[scale=0.38]{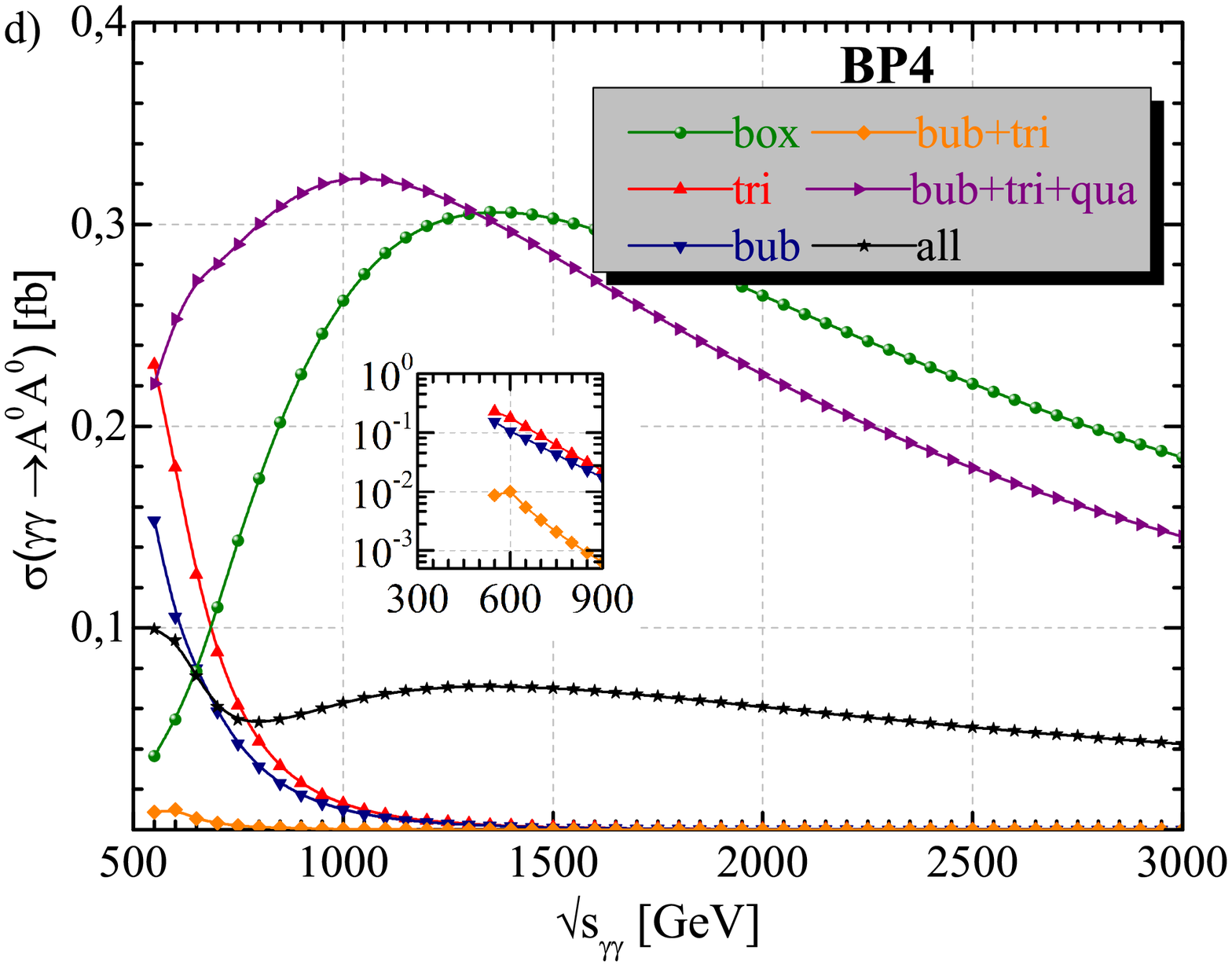}
     \end{center}
\caption{(color online). The individual contributions from each type of diagram to total cross section of process $\gamma \gamma \to A^0 A^0$ as a function of center of mass energy for each benchmark point. The insert figures show the contributions of the triangle-type, buble-type and their interference at the center-of-mass energy range of 300-900 GeV.}
\label{fig:fig4}
\end{figure*}
In Figure~\ref{fig:fig4}, the contribution of each type of diagram to the total cross section is shown as a function of center-of-mass energy for each benchmark point. The labels  ``box'',``tri'', ``bub'',``qua'' and ``all'' represent to the box-type contribution (b$_{1\rightarrow12}$), triangle-type contribution (t$_{1\rightarrow4}$), bubble-type contribution (q$_{1\rightarrow 6}$), quartic-type contribution (q$_{7\rightarrow14}$), and all diagrams contribution, respectively. Additionally, the ``bub+tri'' corresponds to the contribution resulting from interference of bubble-type with triangle-type diagrams.
For each of BPs, the integrated cross section $\hat{\sigma}(\gamma\gamma\rightarrow A^{0}A^{0})$ is enhanced by the threshold effect when $\sqrt{\hat{s}}_{\gamma\gamma}$ is close to 2 times mass of charged-Higgs $H^\pm$, corresponding to the opening of the production channel $\gamma\gamma\rightarrow H^+H^-$. Note that contributions from bubble-type and triangle-type diagrams, which are also called as s-channel contributions, are suppressed at the high energies owing to the s-channel propagator, however the resonant effects can be also seen in these diagrams because of the intermediated neutral Higgs bosons. Particularly, at low center-of-mass energies, $\hat{\sigma}(\gamma\gamma\rightarrow A^{0}A^{0})$ is dominated by triangle-type diagrams since the couplings $h^0A^0A^0$ and $h^0H^+H^-$ are large. At high center-of-mass energy, where the triangle-type contributions are suppressed, $\hat{\sigma}(\gamma\gamma\rightarrow A^{0}A^{0})$ is dominated by the box-type contributions.
However, the bubble-type and triangle-type contributions are almost equal, and their interference (bub+tri) make a much smaller contribution compared to each of them (by one-two orders of magnitude) for
all BPs because they nearly destroy each other. This means that they perform a destructive
interference. Additionally, the box-type contribution is larger than the interference contribution of (bub+tri). Though the quartic-type interactions provide a positive contribution to the total cross section, the sum of the diagrams in quartic-type, bubble-type, and  triangle-type (bub+tri+qua) makes still a small contribution than the box-type diagrams.

The size of $\hat{\sigma}(\gamma\gamma\rightarrow A^{0}A^{0})$ is at a visible level of $10^{-1}$ fb for selected BPs. Furthermore, it is sorted according to BPs as $\hat{\sigma}$(BP1)$>\hat{\sigma}$(BP2)$>\hat{\sigma}$(BP3)$>\hat{\sigma}$(BP4). The basic size of the total cross section differs by few orders of magnitude, depending on the triple and quartic couplings of the pseudoscalar Higgs boson produced.

\begin{figure*}[!t]
    \begin{center}
\includegraphics[scale=0.40]{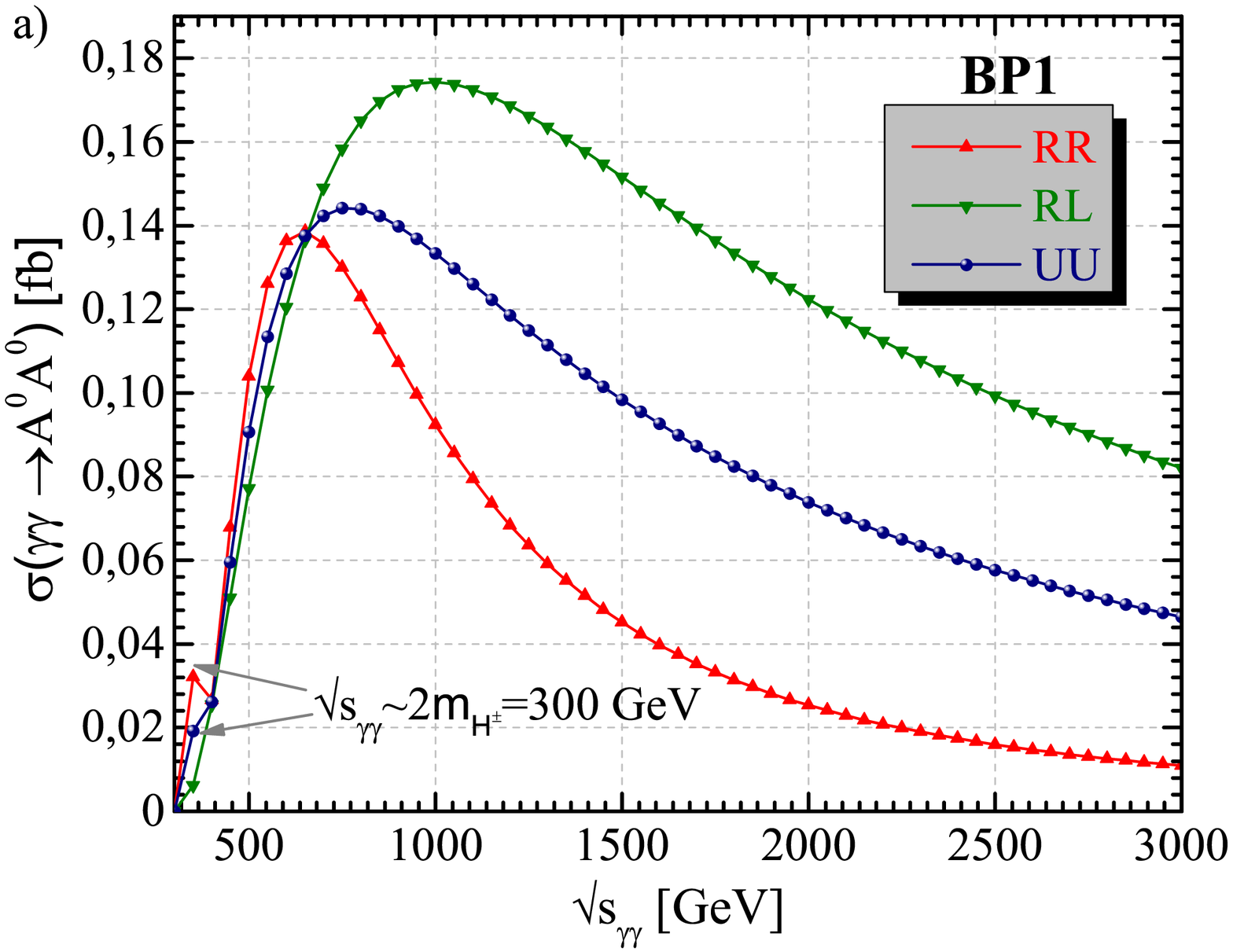}
\includegraphics[scale=0.40]{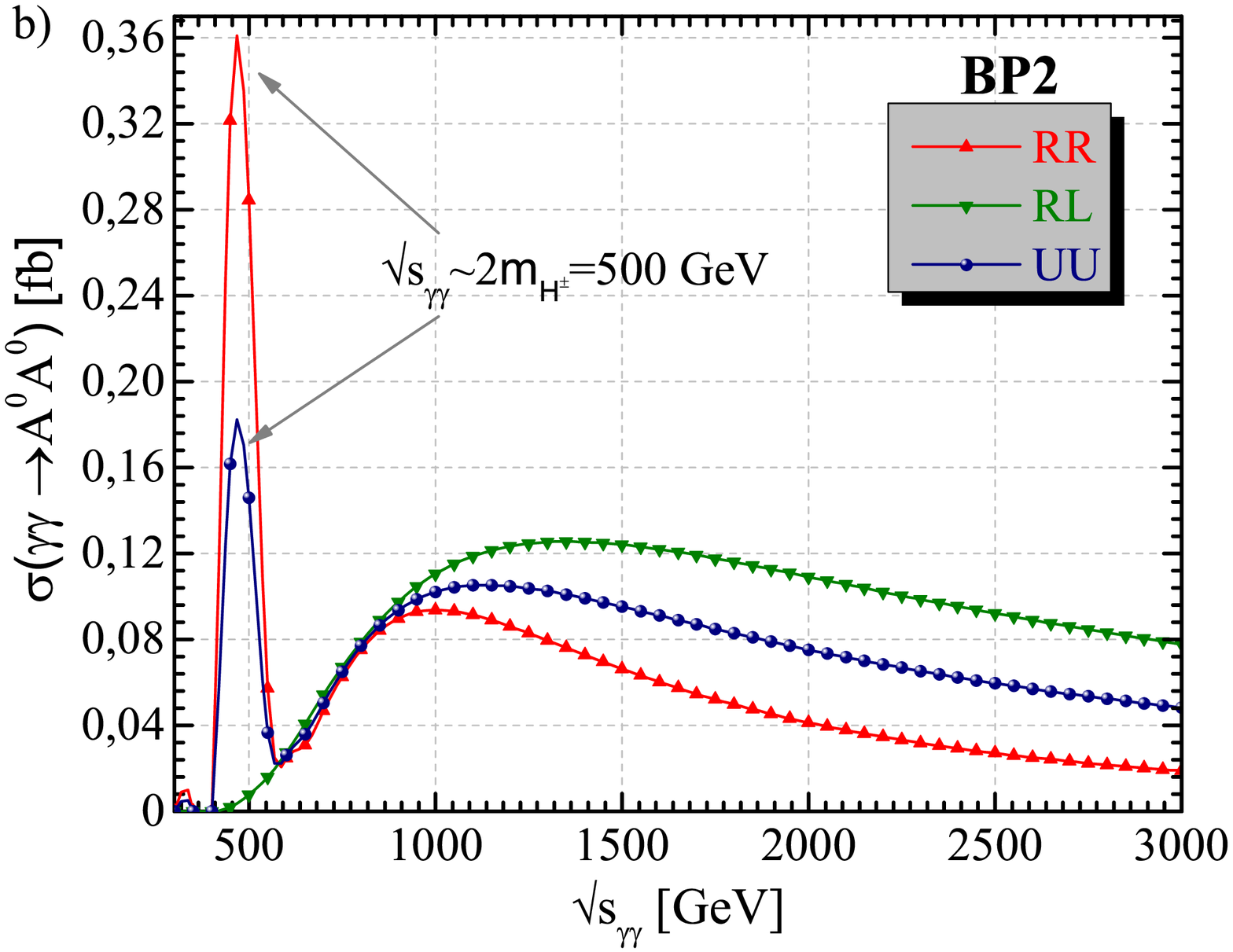}
\includegraphics[scale=0.40]{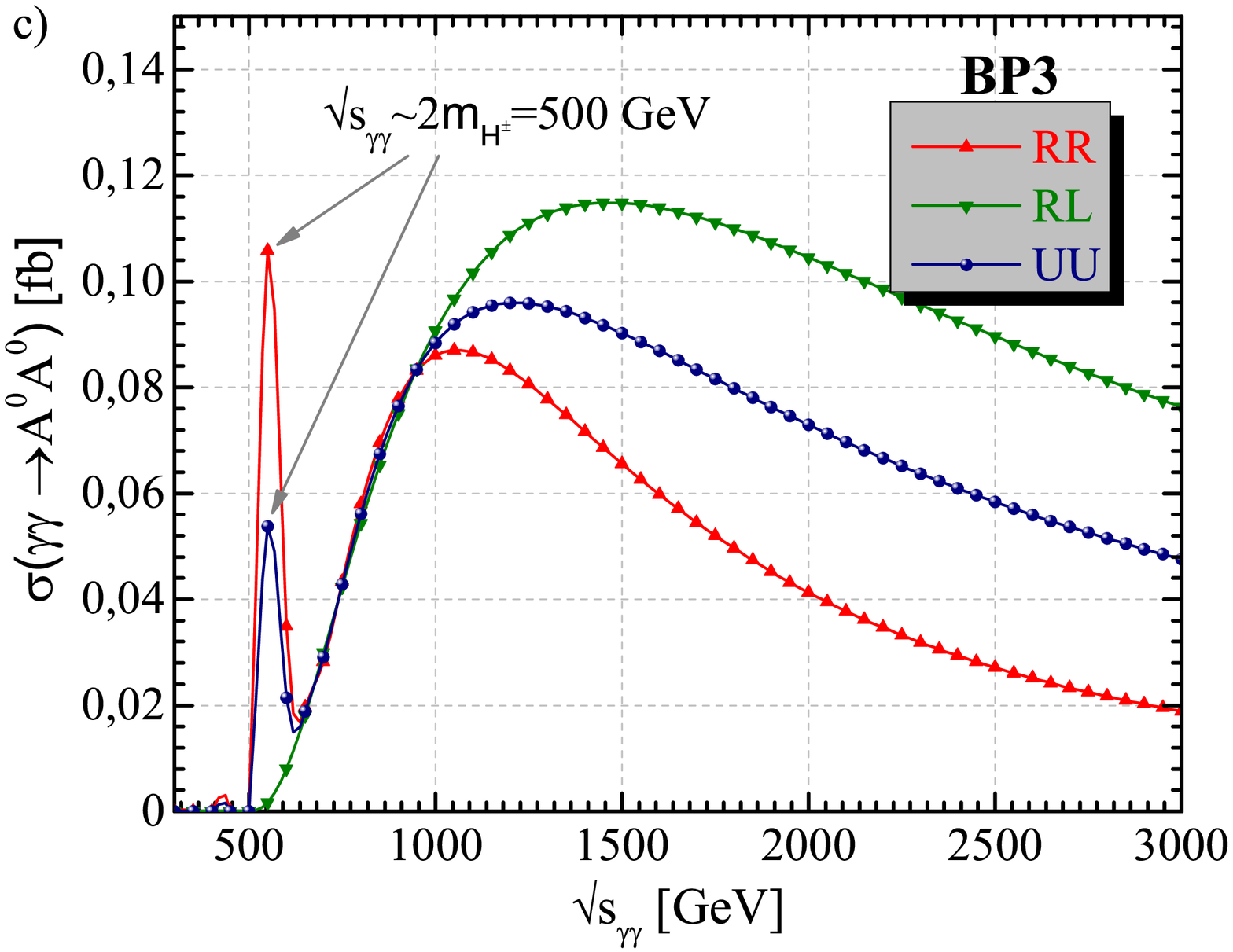}
\includegraphics[scale=0.40]{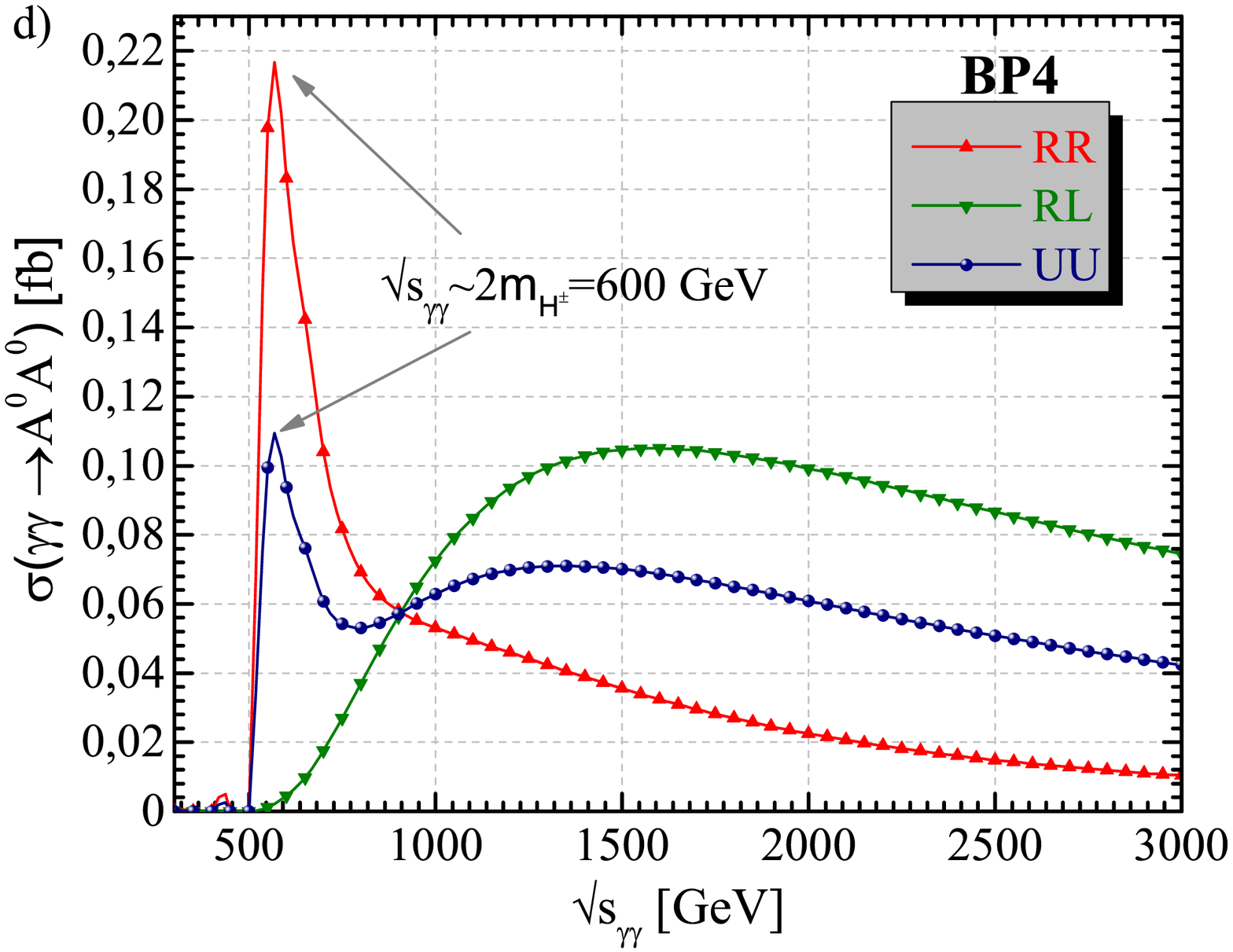}
     \end{center}
\caption{(color online). The total cross section of process $\gamma \gamma \to A^0 A^0$ as a function of center of mass energy for each benchmark point. Also, the threshold effects are shown with arrows. The UU, RR and RL indicate to situations of two photons with unpolarization, right polarization and opposite polarization, respectively.}
\label{fig:fig5}
\end{figure*}
In Figure~\ref{fig:fig5}, the integrated cross section of process $\gamma \gamma \to A^0 A^0$ is given for various polarization configurations of the incoming photons, which are both right-handed $RR$ polarized and opposite polarization $RL$. Note that cross sections are equal in the case of the following polarizations: $\hat{\sigma}(RR)=\hat{\sigma}(LL)$ and $\hat{\sigma}(RL)=\hat{\sigma}(LR)$. It is seen that the unpolarized cross section reaches up to 0.14 fb in BP1 at $\sqrt{\hat{s}}_{\gamma\gamma}=750$ GeV, 0.11 fb in BP2 at $\sqrt{\hat{s}}_{\gamma\gamma}=1150$ GeV, 0.096 fb in BP3 at $\sqrt{\hat{s}}_{\gamma\gamma}=1200$ GeV and 0.07 fb in BP4 at $\sqrt{\hat{s}}_{\gamma\gamma}=1350$ GeV, respectively, and then it falls for all cases. For high center-of-mass energy where the threshold effect are disappeared, when the initial photons have opposite polarizations ($LR$ or $RL$), the total cross section is amplified by a two factor as compared to the unpolarized case ($UU$).
The threshold effect is observed in the case of two photons with left-handed $(LL)$ or right-handed $(RR)$ polarized but not in the case of opposite polarization $(LR)$ or $(RL)$.

\begin{figure}[!h]
    \begin{center}
\includegraphics[scale=0.47]{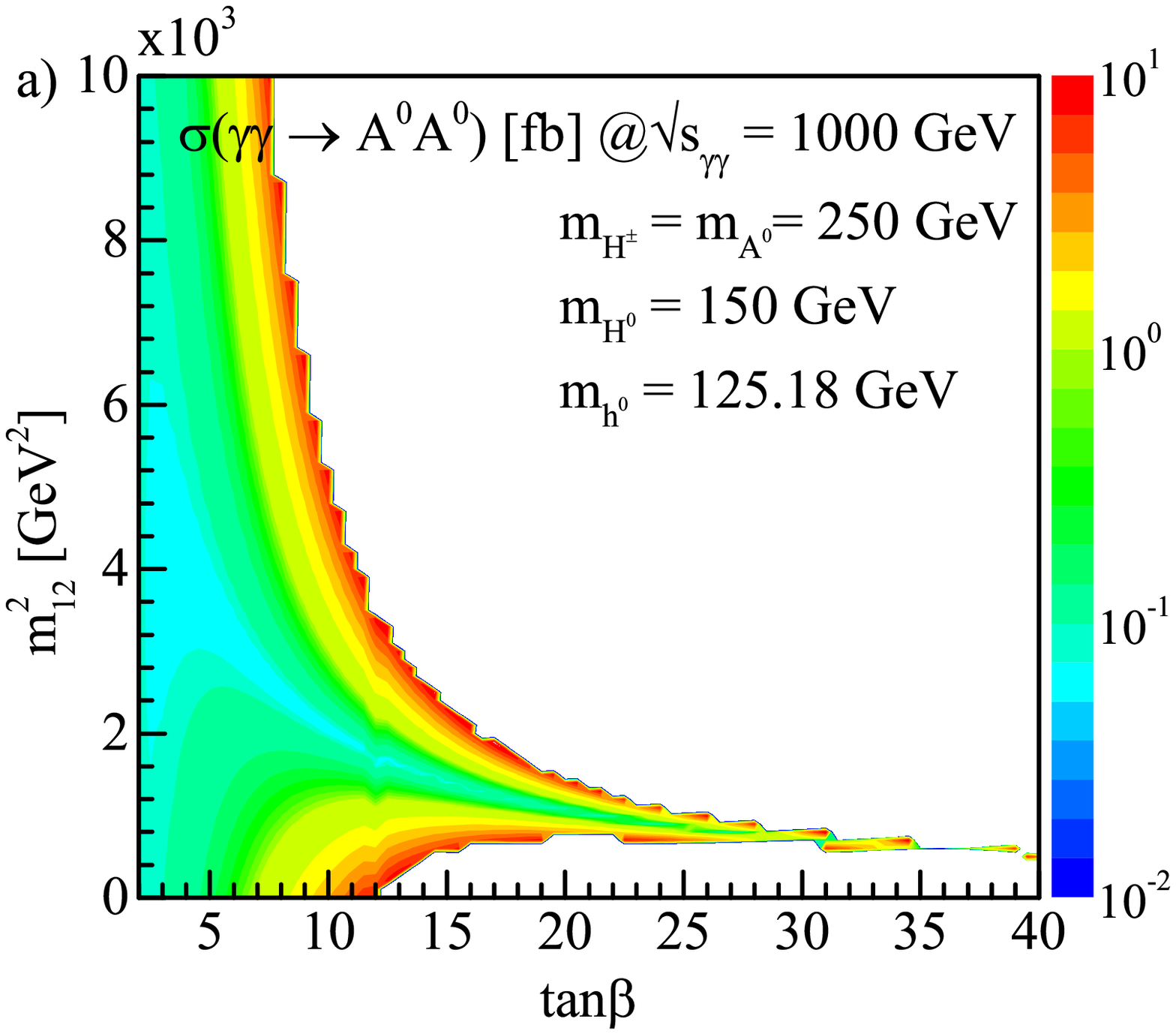}
\includegraphics[scale=0.47]{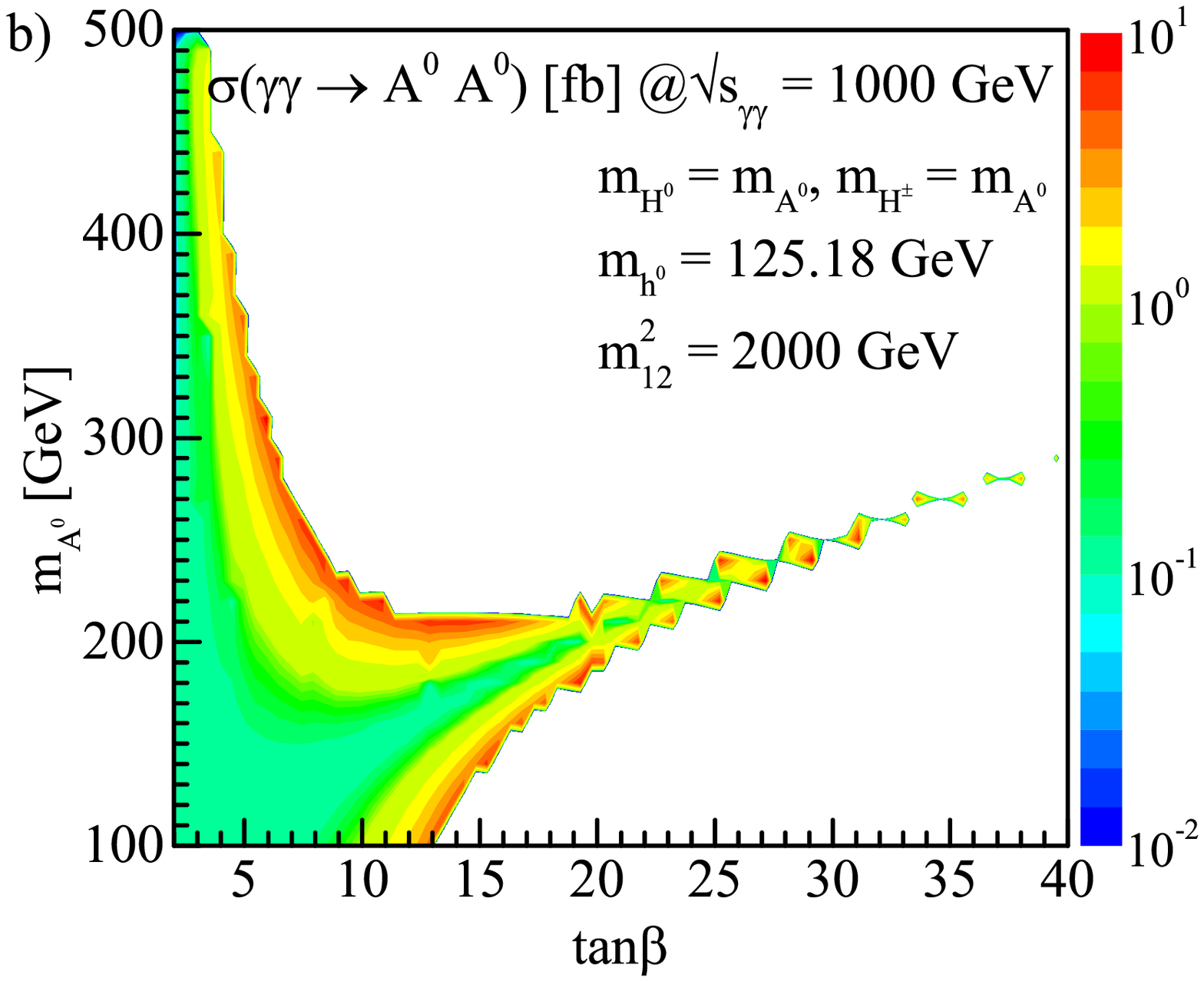}
     \end{center}
\caption{(color online). The total cross section of process $\gamma \gamma \to A^0 A^0$ as a 2D function of parameters a) $m_{12}^2-\tan \beta$ and b) $m_A-\tan \beta$ for $\sqrt{\hat{s}}=1$  TeV. The colour heat map corresponds to the total cross section (in fb) in the scan region. The white region represents
parameter space where production of $A^0 A^0$
is kinematically unavailable as well as not allowed by theoretical constraints.}
\label{fig:fig6}
\end{figure}
It is well known that the triple and quartic couplings of the pseudoscalar Higgs boson to other particles depend on the soft breaking parameter $m_{12}^2$, pseudoscalar Higgs boson mass and $\tan\beta$. The dependence of the cross section on these parameters can provide important information about these couplings. From this dependence one can revealed a region of the parameter space where the enhancement of cross section is large enough to be detectable at future linear colliders. In respect to this, the integrated cross section of $\gamma \gamma \to A^0 A^0$ is scanned over the regions of $m_{12}^2-\tan \beta$ and $m_A-\tan \beta$ at $\sqrt{\hat{s}}=1$  TeV, as shown in Figs.~\ref{fig:fig6}(a) and ~\ref{fig:fig6}(b).
The scan parameters are varied as follows $0\leq m_{12}^2\leq 10^4$ GeV$^2$ in steps of 100 GeV, $100\leq m_{A}\leq 500$ GeV in steps of 10 GeV, and $2\leq\tan \beta \leq 40$ in steps of 0.5. However, most of parameter region is reduced by theoretical constraints as well as due to kinematically inaccessible. The cross section is enhanced at the border of the allowed and not-allowed regions. The size of $\sigma(\gamma\gamma\rightarrow A^{0}A^{0})$ is at a visible level of $10^{-1}$ to $10^{1}$ fb for considered parameter regions.

\begin{figure}[!t]
    \begin{center}
\includegraphics[scale=0.42]{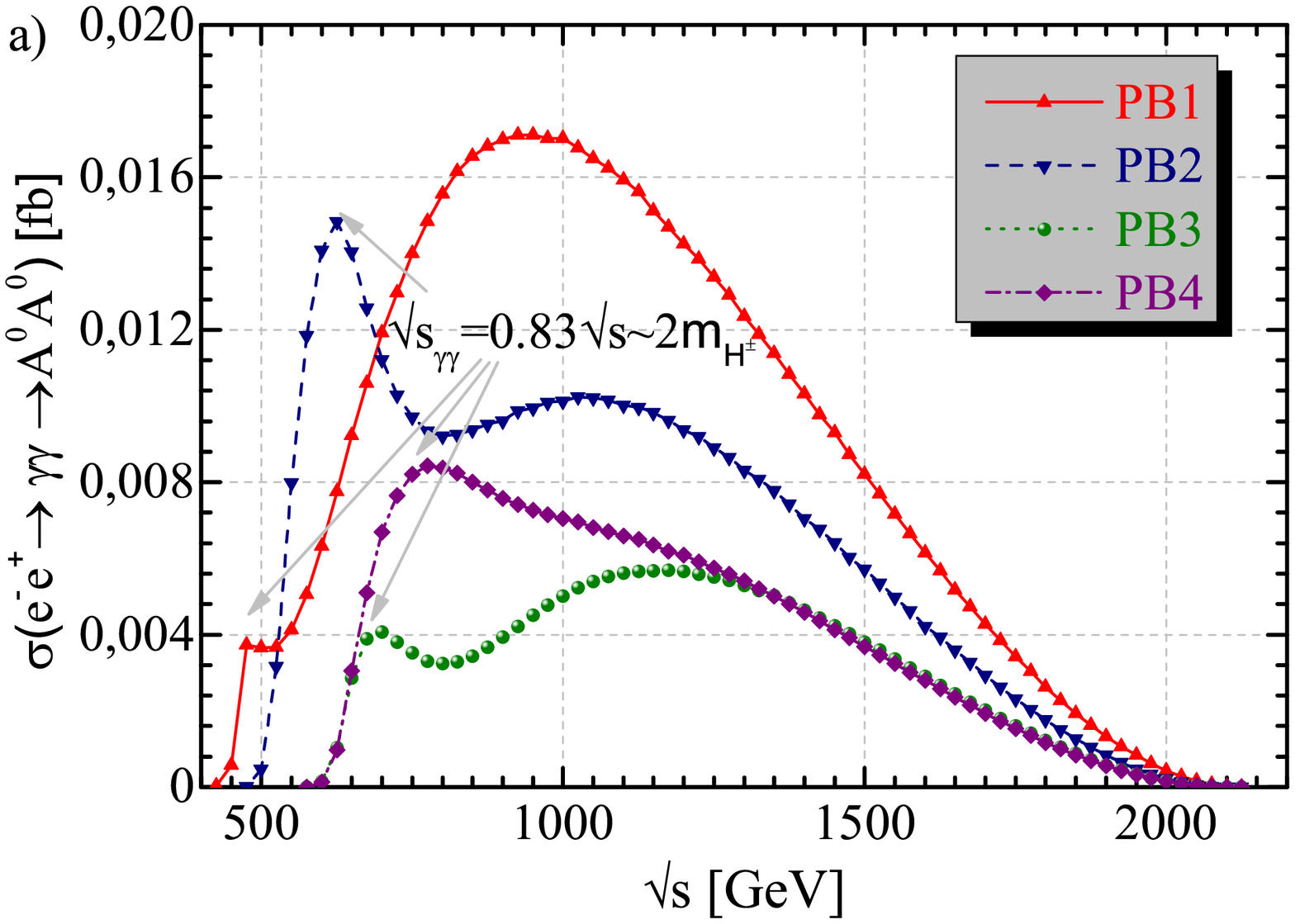}
\includegraphics[scale=0.43]{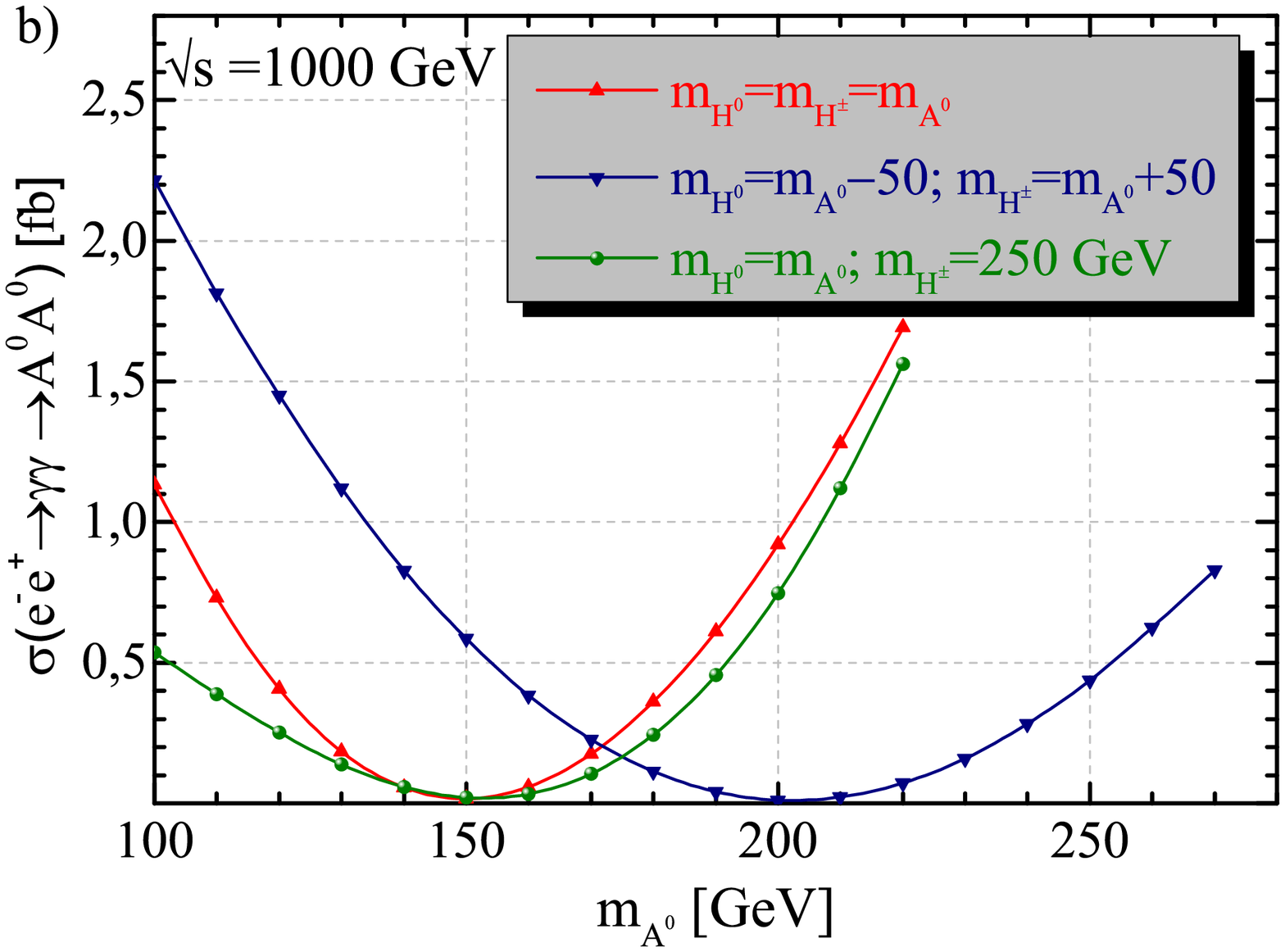}
     \end{center}
\caption{(color online). The total cross section of process $e^{-}e^{+} \to \gamma \gamma \to A^0 A^0$ as a function of (a) center of mass energy for each benchmark point and (b) pseudoscalar Higgs mass for different Higgs mass hierarchy.}
\label{fig:fig7}
\end{figure}
Finally, Figure~\ref{fig:fig7} shows the total cross section of process $e^{-}e^{+} \to \gamma \gamma \to A^0 A^0$ as a function of center of mass energy for each BPs and pseudoscalar Higgs mass for several Higgs mass hierarchy. $\sigma(e^{-}e^{+} \to\gamma\gamma\rightarrow A^{0}A^{0})$  is evaluated by convoluting $\sigma(\gamma\gamma\rightarrow A^{0}A^{0})$ with the photon luminosity spectrum based Compton backscattered photons. It can be easily seen that the total cross section in the BP1 scenario is larger than others. The total cross section is enhanced by the threshold effect when $0.83\sqrt{s}\sim 2m_{H^\pm}$, corresponding to the opening of the charged Higgs pair channel $\gamma\gamma\rightarrow H^+H^-$.
The size of $\sigma(e^{-}e^{+} \to \gamma\gamma\rightarrow A^{0}A^{0})$ is at a visible level of $10^{-2}-10^{-3}$ fb, depending on BPs. For selected BPs, the pseudoscalar Higgs boson pair production is more likely to be observed in the $\gamma \gamma$ collider than in the electron-positron collider. Total cross section decreases with increasing pseudoscalar Higgs mass up to a certain threshold, and then it increases due to the resonant effects. In particular, the total cross section reaches a value of 2.22 fb for $m_{A^0}=100$ GeV in the case of $m_{H^\pm}=m_{H^0}=m_{A^0}$.

\section{Conclusion}\label{sec:conc}
In this study,  nonexistent at tree level, and first appeared at one-loop level, the process $\gamma \gamma \to A^0 A^0$ have been studied with special emphasis put on individual contributions from each type diagram at a $\gamma \gamma$ collider as well as electron-positron collider. The calculation was carried out in the framework of THDM taking into account both theoretical restrictions and experimental constraints from recent LHC data and other experimental results.
Energy-dependent structure of the cross section is revealed by resonance effects due to the intermediate neutral Higgs boson as well as by the threshold effect when $\sqrt{s}_{\gamma\gamma}\sim 2m_{H^\pm}$.  For all cases, the box-type diagrams make dominant contribution at high energies. Owing to the CP nature of $A^0$, the box-type diagrams are rather sensitive to the coupling $\lambda_{H^\pm G^\mp A^0}$ which does not have neither a $\tan\beta$ nor a $m_{12}^2$ dependence. Hence, the process $\gamma \gamma \to A^0 A^0$ makes it possible to detect the coupling $\lambda_{H^\pm G^\mp A^0}$. However, triple couplings $\lambda_{[h^0,H^0] A^0 A^0 }$, $\lambda_{H^+ H^-[h^0,H^0]}$ and $\lambda_{G^+ G^-[h^0,H^0]}$ only appear at s-channel diagrams and they can be amplified by resonance effects due to neutral Higgs bosons. The polarization configurations of the initial photons have the potential to amplify the total cross sections which can develop the number of events to be detected at the future colliders. Consequently, the cross section of pair production of the pseudoscalar Higgs boson at a photon-photon collision could be considerably amplified in the THDM and, therefore, the number of events expected at a $\gamma \gamma$ collider can allow you to determine or exclude the parameter space of the THDM potential.


\end{document}